\documentclass[reprint,
aps,prb,
10pt,
nofootinbib
]{revtex4-2}

\usepackage{microtype}

\usepackage{amsmath}
\usepackage{amssymb}
\usepackage{graphicx}
\usepackage{hyperref}
\usepackage{dsfont}
\usepackage{newtxtext}
\usepackage[varvw]{newtxmath}
\usepackage{slashed}
\usepackage{bbm}
\usepackage{tabularx}

\hypersetup{
	colorlinks,
	linkcolor={blue},
	citecolor={blue},
	urlcolor={blue}
}

\graphicspath{{./}{./figs/}}

\newcommand{\rme}{\mathrm{e}}
\newcommand{\rmi}{\mathrm{i}}
\newcommand{\rmd}{\mathrm{d}}
\newcommand{\Nb}{{N_\text{b}}}
\newcommand{\Nf}{N_\text{f}}
\newcommand{\SUtwo}{\text{SU(2)}}

\renewcommand{\vec}[1]{\boldsymbol{#1}}

\usepackage[dvipsnames]{xcolor}
\usepackage[normalem]{ulem}

\begin{document}
	
\title{%
Gross-Neveu-Heisenberg criticality from competing nematic and antiferromagnetic orders in bilayer graphene
}
	
\author{Shouryya Ray}
\author{Lukas Janssen}
	
\affiliation{Institut f\"ur Theoretische Physik and W\"urzburg-Dresden Cluster of Excellence ct.qmat, TU Dresden, 01062 Dresden, Germany}
	
\begin{abstract}
We study the phase diagram of an effective model of competing nematic and antiferromagnetic orders of interacting electrons on the Bernal-stacked honeycomb bilayer, as relevant for bilayer graphene. In the noninteracting limit, the model features a semimetallic ground state with quadratic band touching points at the Fermi level. Taking the effects of short-range interactions into account, we demonstrate the existence of an extended region in the mean-field phase diagram characterized by coexisting nematic and antiferromagnetic orders. By means of a renormalization group approach, we reveal that the quantum phase transition from nematic to coexistent nematic-antiferromagnetic orders is continuous and characterized by emergent Lorentz symmetry. It falls into the $(2+1)$-dimensional relativistic Gross-Neveu-Heisenberg quantum universality class, which has recently been much investigated in the context of interacting Dirac systems in two spatial dimensions. The coexistence-to-antiferromagnetic transition, by contrast, turns out to be weakly first order as a consequence of the absence of the continuous spatial rotational symmetry on the honeycomb bilayer. Implications for experiments in bilayer graphene are discussed.
\end{abstract}

\date{July 1, 2021} 

\maketitle

\section{Introduction}
	
Since its experimental realization~\cite{novoselov06}, the low-temperature physics of bilayer graphene has attracted significant attention.
However, despite considerable experimental and theoretical efforts, the actual nature of the material's zero-temperature ground state has not been unambiguously identified to date.
The problem is that many competing states, which are very close in energy, appear in the system~\cite{jung11}. Already slight changes in the experimental setup or conditions can therefore lead to qualitatively different low-temperature ground states.
In the simplest tight-binding model for the thermodynamically stable Bernal-stacked configuration of bilayer graphene~\cite{mccann06}, the valence and conduction bands touch quadratically at two isolated points in the Brillouin zone, with the Fermi level being locked at the band touching point in the undoped system.
In contrast to monolayer graphene, the single-particle density of states in the bilayer model hence remains finite at the Fermi level, such that the semimetallic state is prone to instabilities at low temperatures~\cite{sun09, vafek10a, zhang10}. 
Transport and spectroscopic experiments have indeed observed an interaction-driven reconstruction of the fermionic spectrum at temperatures below around 10 K~\cite{feldman09, martin10, weitz10, mayorov11, velasco12, freitag12, bao12, veligura12}. 
However, while some of the experiments indicate an insulating ground state with a full bulk band gap~\cite{bao12, freitag12, veligura12, velasco12}, others suggest only a partial gap opening in which four isolated Dirac cones remain gapless in the bulk spectrum~\cite{mayorov11}.
A partial gap opening would imply a low-temperature ground state that breaks part of the lattice rotational symmetry spontaneously. In fact, such an electronic nematic order had indeed previously been predicted on the basis of perturbative renormalization group (RG) analyses~\cite{vafek10a, lemonik10}.
Later theoretical studies~\cite{vafek10b, cvetkovic12, lemonik12, scherer12, lang12, pujari16, honerkamp17, leaw19} have shown, however, that an antiferromagnetic state, characterized by finite and opposite net magnetizations within the two layers~\cite{kharitonov12, lang12}, is at least comparable in energy and in fact prevails over a large section of parameter space. This layer antiferromagnet features a full gap in the electronic spectrum, and among the different candidate ground states it appears to be the one that is most consistent with the measurements on the samples that become insulating at low temperatures~\cite{velasco12}.
	
\begin{figure}[tb]
	\includegraphics[width=\linewidth]{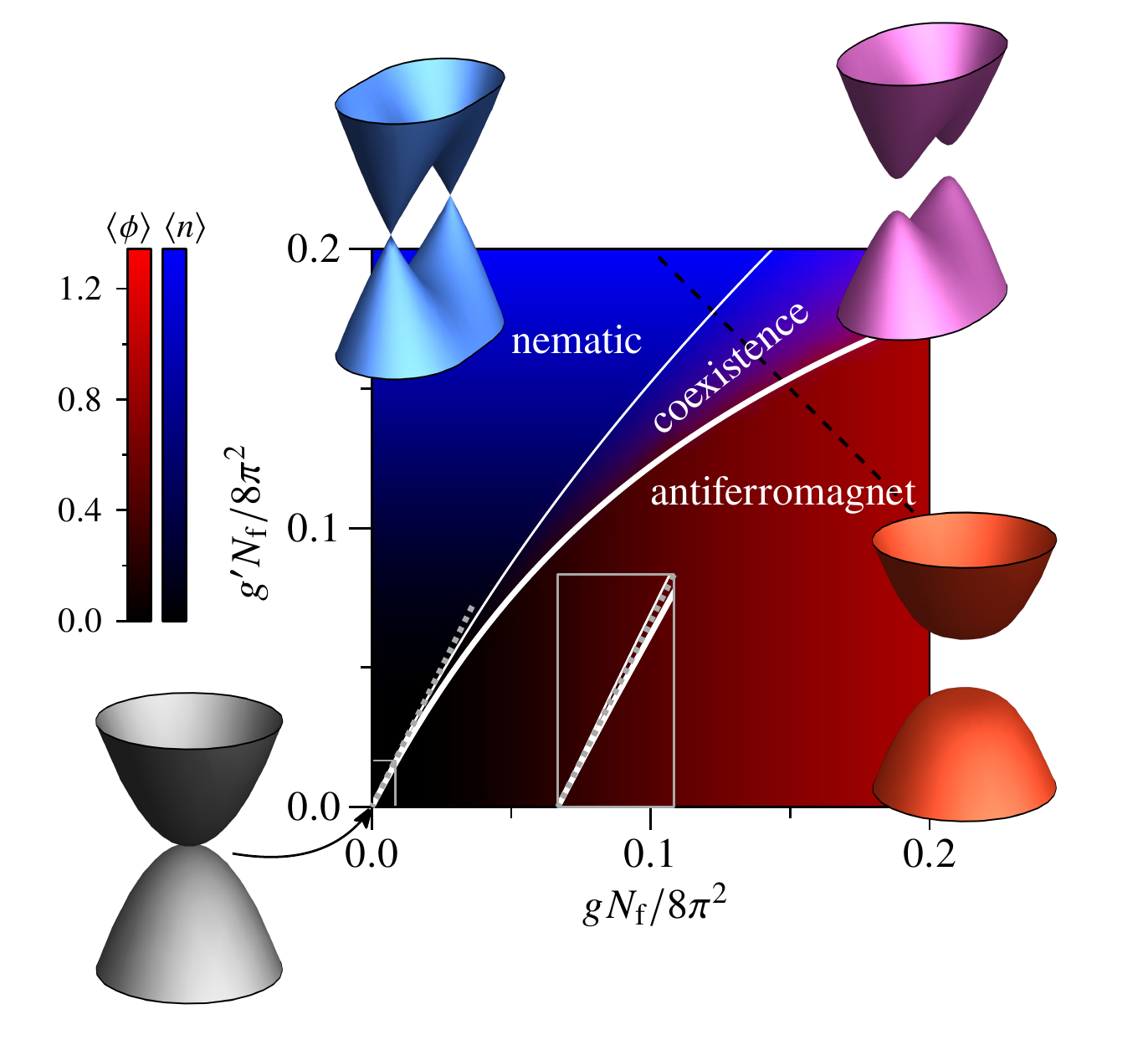}
	\caption{Mean-field phase diagram of interacting electrons on the Bernal-stacked honeycomb bilayer as a function of short-range couplings $g$ and $g'$ defined in Sec.~\ref{sec:model}. Blue and red color codings indicate the magnitudes of the nematic and antiferromagnetic orders, respectively. The electronic spectra near the corners of the hexagonal Brillouin zone are depicted for the different states in the insets. The gray rectangle shows a zoom into the weakly interacting regime, with the dotted gray line indicating the phase boundary between nematic and antiferromagnetic orders in the fermionic RG calculation (Sec.~\ref{sec:phasediagram}). The dashed black line indicates the cut used in Fig.~\ref{fig:order-parameters}. The antiferromagnetic-to-coexistence transition (thick white curve) is weakly first order (Sec.~\ref{sec:afm-coexistence}). The nematic-to-coexistence transition (thin white curve) is continuous and falls into the Gross-Neveu-Heisenberg universality class (Sec.~\ref{sec:nem-coexistence}).}
	\label{fig:MFTphasediag}
\end{figure}

In this work, we revisit the problem of the low-temperature ground state in bilayer graphene. We investigate the phase diagram of a model of short-range-interacting electrons on the honeycomb bilayer by means of mean-field and RG analyses.
We focus on the competition between the nematic and antiferromagnetic orders, which appear to be the most promising candidate ground states consistent with the experiments~\cite{mayorov11, velasco12}. In particular, we study the possibility of coexisting orders, which was only insufficiently addressed in previous work~\cite{cvetkovic12}.
We find that the nematic and antiferromagnetic phases at small to moderate coupling are generically separated by an intermediate coexistence phase which features both layer antiferromagnetism and nematicity, see Fig.~\ref{fig:MFTphasediag}.
In the coexistence phase, the fermionic spectrum exhibits a full but rotationally anisotropic band gap.
We also discuss the quantum transitions in and out of the coexistence phase, the nature of which may have nontrivial consequences for various experimental observables.
The fact that the rotational symmetry on the honeycomb bilayer is restricted to 120$^\circ$ rotations allows a cubic invariant in the effective potential for the nematic order parameter. We show that the presence of this term renders the transition between the gapped antiferromagnet and the coexistence phase weakly first order.
By contrast, we find the transition from the nematic phase to the coexistence phase to be continuous as a consequence of an emergent relativistic space-time symmetry realized at large length scales. 
At this transition, the electronic band gap closes at four isolated Fermi points in the Brillouin zone with linear band dispersions in their vicinities. 
We characterize this quantum critical point within two complementary approaches: (i) a $2+\varepsilon$ expansion around the lower critical space-time dimension of two within the purely fermionic theory, and (ii) a $4-\epsilon$ expansion around the upper critical space-time dimension of four within the corresponding Hubbard-Stratonovich-decoupled fermion-boson model.
The consistent result of the two approaches is that the nematic-to-coexistence transition falls into the relativistic Gross-Neveu-Heisenberg universality class.
This quantum universality class has recently been much investigated in models of interacting electrons on monolayer honeycomb or $\pi$-flux lattices~\cite{%
	herbut06, herbut09b, janssen14, zerf17, knorr18,
	gracey18,
	assaad13, toldin15, otsuka16, otsuka20, buividovich18, lang19, liu19, liu21, xu21}.
It was also proposed in the context of the Hubbard model on the Bernal-stacked honeycomb bilayer, in which a fully symmetric Dirac semimetal phase is stabilized in the weakly interacting regime as a consequence of a trigonal warping term arising from fermion self-energy effects~\cite{pujari16, ray18}.
The mechanism proposed here is different from these latter works in that it applies to the ordered regime, in which the role of the interaction-induced trigonal warping becomes subdominant~\cite{cvetkovic12, honerkamp17, hesselmann20}.
The background nematic order present throughout this transition gaps out half of the Dirac cones of the symmetric Dirac semimetal state, leading to a smaller number of low-energy fermion degrees of freedom and critical power laws with different exponents.
The exponents are also different from those of the Hubbard model of spin-$1/2$ fermions on the monolayer.
As each quadratic band touching point in the noninteracting fermion spectrum splits into two mini-Dirac cones in the nematic state, the number of fermion degrees of freedom is doubled in comparison with the monolayer case.
As an aside, we note that the nontrivial quantum transitions we find requires that the model features the full symmetries of bilayer graphene on the microscopic level, with the spatial and/or spin rotational symmetry being broken only spontaneously. This is in contrast to previous work on the coexistence of nematic and gapped states in bilayer graphene, which assumed explicit breaking of microscopic symmetries, e.g., by external strain and/or gate voltage~\cite{gorbar12}.

The remainder of the paper is organized as follows: In Sec.~\ref{sec:model}, we describe our model of interacting electrons on the honeycomb bilayer. Section~\ref{sec:phasediagram} discusses the mean-field phase diagram for the competing nematic and antiferromagnet orders. Our results for the antiferromagnet-to-coexistence and nematic-to-coexistence transitions are presented in Secs.~\ref{sec:afm-coexistence} and \ref{sec:nem-coexistence}, respectively. We conclude in Sec.~\ref{sec:conclusions}. Technical details are deferred to three appendices.

\section{Model}
\label{sec:model}
	
Since a fully satisfactory microscopic model of the electronic interactions in bilayer graphene is currently not agreed upon,%
\footnote{See, nevertheless, Ref.~\cite{wehling11} for ab-initio results for monolayer graphene and bulk graphite, as well as Ref.~\cite{zhang08} for an overview of band structure model parameters for bilayer graphene.}
we employ in this work a minimal theoretical description that allows us to study the competition between nematic and antiferromagnetic orders and the possibility of a coexistence phase on the honeycomb bilayer.
Our approach may be viewed as a simple phenomenological modeling that captures the physics of two most prominent candidate ordered states discussed in the experimental works~\cite{mayorov11, velasco12}. It restricts the multidimensional parameter space discussed in previous more comprehensive works~\cite{cvetkovic12, lemonik12, szabo21} to a simple two-dimensional subspace.
Explicitly, we consider the low-energy continuum action $S = \int \mathrm d\tau \mathrm d^2 \vec{x} \mathcal L_\text{QBT}$ in imaginary time $\tau$ and two-dimensional space $\vec x = (x,y)^\top$ with
\begin{align}
	\mathcal{L}_\text{QBT} &= \Psi^{\dagger}[\partial_\tau + d_a(-\rmi \nabla) (\Gamma_a \otimes \mathds{1}_2)]\Psi \nonumber\\
	&\phantom{{}={}} - \frac{g}{2}\!\left[ \Psi^\dagger (\Gamma_z \otimes \sigma_\alpha)\,\Psi \right]^2 - \frac{g'}{2}\!\left[ \Psi^\dagger (\Gamma_a \otimes \mathds{1}_2) \Psi \right]^2,
	\label{eq:phenoL}
\end{align}
where $a=1,2$ and $\alpha = x,y,z$.
In the above and the following equations, the summation convention over repeated indices is implicitly assumed. The $d_a$ functions are $\ell = 2$ real spherical harmonics given by
\begin{align}
	d_1(-\rmi \nabla) = -\partial_x^2 + \partial_y^2, \qquad 
	d_2(-\rmi \nabla) = -2\partial_x \partial_y,
\end{align}
and transform under spatial rotations as components of a second-rank tensor \cite{janssen15}.
The spinors $\Psi$ and $\Psi^\dagger$ have eight components, corresponding to the layer, valley, and physical spin degrees of freedom \cite{vafek10a}. The $2\times 2$ Pauli matrices $\sigma_\alpha$, $\alpha = x,y,z$, act on the physical spin index and transform as a vector under $\SUtwo$ spin rotations.
The $4\times 4$ matrices $\Gamma_x$, $\Gamma_y$, and $\Gamma_z$ realize a four-dimensional representation of the Clifford algebra, and are given explicitly as
\begin{align}
	\Gamma_x = \mathds{1}_2 \otimes \mu_x, \quad \Gamma_y = \tau_z \otimes \mu_y, \quad \Gamma_z = \mathds{1}_2 \otimes \mu_z, \label{eq:matsGamma}
\end{align}
where in the above tensor products the first (second) factors act on the layer (valley) indices.
Here, the $2\times 2$ Pauli matrices that serve as building blocks for the $\Gamma$ matrices have been denoted by $\tau_\alpha$ and $\mu_\alpha$ to distinguish them from those acting on the physical spin index. 
While $\Gamma_x$ and $\Gamma_y$ transform as components of a second-rank tensor, $\Gamma_z$ is a scalar under spatial rotations~\cite{ray18}.
In this representation, the time-reversal operator is given as $\mathcal{T} = (\tau_x \otimes \mathds{1}_2) \otimes \sigma_y\,\mathcal{K}$, where $\mathcal{K}$ denotes complex conjugation. The first factor of the unitary part essentially represents interchanging the two valleys, while the second factor represents spin flip.
In our model~\eqref{eq:phenoL}, we have assumed particle-hole symmetry and a continuous spatial rotational symmetry. The effects of perturbations that break the continuous rotational symmetry down to 120$^\circ$ rotations on the honeycomb bilayer will be discussed later in the paper.
In particular, our model neglects the effects of trigonal warping that are expected to play a dominant role only in the weakly interacting regime~\cite{cvetkovic12, pujari16, honerkamp17, ray18, hesselmann20}.
We use units in which the isotropic effective band mass is set to $m^* = 1/2$. 
The spectrum of the noninteracting Hamiltonian $\mathcal H_0(\vec p) = d_a(\vec p) (\Gamma_a \otimes \mathds{1}_2)$ then is simply $\varepsilon^\pm_0(\vec p) = \pm \vec p^2$, where $\vec p$ denotes the deviation from the corners $\vec K$ and $\vec K'$ of the hexagonal Brillouin zone. It describes a nonrelativistic two-dimensional ``Luttinger'' semimetal~\cite{ray20} in which the valence and conduction bands touch quadratically at the two Fermi points at $\vec K$ and $\vec K'$.

The four-fermion interactions parametrized by the couplings $g$ and $g'$ in Eq.~\eqref{eq:phenoL} are chosen such that they stabilize antiferromagnetic and nematic long-range order, respectively.
This can be seen as follows:
The three-component fermion bilinear $\vec\phi \sim \Psi^\dagger (\Gamma_z \otimes \vec{\sigma}) \Psi$, associated at the mean-field level with the four-fermion coupling $g$, is even under time reversal, a scalar under spatial rotations, and a vector under $\SUtwo$ spin rotations. Assigning a finite vacuum expectation value to $\vec\phi$ hence breaks spin-rotational symmetry while leaving spatial rotational symmetry and time reversal intact. It describes the layer antiferromagnet, in which the two honeycomb layers feature finite and opposite magnetizations~\cite{cvetkovic12, kharitonov12}. Importantly, the corresponding operator $\Gamma_z \otimes \vec\sigma$ anticommutes with the single-particle Hamiltonian $\mathcal H_0$, and hence a vacuum expectation value of $\vec\phi$ opens a uniform gap in the fermionic spectrum, of size $\propto | \langle \vec\phi \rangle |$.
Microscopically, the four-fermion term parametrized by $g$ can be understood to arise from an interlayer interaction that couples spin densities on the two honeycomb layers~\cite{vafek10b, cvetkovic12}.
On the other hand, the bilinear corresponding to the coupling $g'$, $n_a \sim \Psi^\dagger (\Gamma_a \otimes \mathds{1}_2) \Psi$, transforms as the components of a second-rank tensor under spatial rotations, while being even under spin rotations and time reversal. When $n_a$ obtains a finite expectation value, the spatial rotational symmetry on the honeycomb bilayer is spontaneously broken while all other symmetries are left intact. The bilinear $n_a$ corresponds to the nematic order parameter~\cite{vafek10a, cvetkovic12}. Its components commute with one of the matrices appearing in the single-particle Hamiltonian, while anticommuting with the other. A gap in the electronic spectrum is therefore not opened up in the state with nematic order alone; instead, each quadratic band touching point splits into two mini-Dirac cones located in close vicinity of the corners $\vec K$ and $\vec K'$ of the hexagonal Brillouin zone, along the axis chosen by $\langle n_a \rangle$.
On a microscopic level, the coupling $g'$ can be thought of as parametrizing intervalley scattering processes between the $\vec K$ and $\vec K'$ points~\cite{vafek10a, vafek10b, cvetkovic12}.

The symmetry of the noninteracting Hamiltonian allows a number of further short-range interactions~\cite{vafek10b}, which are neglected here for simplicity. These may change some of our results on a quantitative level, such as the size of the phases and the location of the phase boundaries in parameter space. However, our main conclusions, including the existence of a coexistence phase and the nature of the transitions into and out of this phase are expected to be robust upon the inclusion of these further interactions, as long as these become not too large.
The same holds for the long-range tail of the Coulomb interaction, which may be included as well, but is expected to be screened at low energies~\cite{lemonik12}.

\section{Phase diagram}
\label{sec:phasediagram}

In this section, we explore the phase diagram of the model~\eqref{eq:phenoL} as a function of the coupling parameters $g$ and $g'$ on the level of mean-field theory. We restrict ourselves to positive interactions $g,g' > 0$, which allows us to obtain an equivalent order-parameter field theory by means of a  Hubbard-Stratonovich transform,
\begin{align}
	\mathcal{L}_\text{HST} &= \frac{\phi^2}{2g} + \frac{n^2}{2g'} + \Psi^\dagger\left[\partial_\tau + d_a(-\rmi \nabla) (\Gamma_a \otimes \mathds{1}_2)\right]\Psi
	\nonumber\\ & \quad 
	- \phi_\alpha \Psi^\dagger(\Gamma_z \otimes \sigma_\alpha)\Psi - n_a \Psi^\dagger(\Gamma_a \otimes \mathds{1}_2)\Psi,
\end{align}
where $\phi^2 \equiv \phi_\alpha \phi_\alpha$, $\alpha = x,y,z$, and $n^2 \equiv n_a n_a$, $a = 1,2$. The collective fields $\phi_\alpha$ and $n_a$ are related to fermion bilinears via the equations of motion $\phi_\alpha = g\Psi^\dagger(\Gamma_z \otimes \sigma_\alpha)\Psi$ and $n_a = \Psi^\dagger (\Gamma_a \otimes \mathds{1}_2) \Psi$.
We integrate out the fermions by performing the path integral of $\Psi$ and $\Psi^\dagger$ in $\mathcal L_\text{HST}$ in order to obtain an effective description in terms of the two order parameters alone,
\begin{align}
	\mathcal L_\text{OP} &= \frac{\phi^2}{2g} + \frac{n^2}{2g'} - \frac{\Nf}{4} \operatorname{Tr} \ln \big[\partial_\tau + d_a(-\rmi \nabla) (\Gamma_a \otimes \mathds{1}_2) 
	\nonumber\\ & \quad 
	- \phi_\alpha (\Gamma_z \otimes \sigma_\alpha) - n_a (\Gamma_a \otimes \mathds{1}_2)\big].
\end{align}
In the above, we have inserted a parameter $\Nf$ which counts the number of valley and spin degrees of freedom, with $\Nf = 4$ corresponding to the present case of spin-$1/2$ fermions on the honeycomb bilayer.
In the limit $\Nf \to \infty$, bosonic fluctuations freeze out and mean-field theory becomes exact. 
We can then replace $\phi_\alpha$, $n_a$ with corresponding classical fields and perform the trace in momentum space.
Evaluating the frequency integral, we find the familiar sum over energy of filled states for the mean-field effective potential
\begin{align}
	V_{\text{MF}}(\phi,n_a) = \frac{\phi^2}{2g} + \frac{n^2}{2g'} + \Nf \int_{|\vec{p}| \leqslant \Lambda} \frac{\rmd^2 \vec{p}}{(2\pi)^2}\,\varepsilon^-_{\phi,n}(\vec{p}),
	\label{eq:VeffMFmaster}
\end{align}
where
\begin{align}
	\varepsilon^\pm_{\phi,n}(\vec{p}) = \pm\sqrt{p^4 + \phi^2 + n^2 - 2n_a d_a(\vec{p})}
\end{align}
denotes the fermion spectrum in the presence of a constant bosonic background, and $\Lambda$ is a ultraviolet momentum cutoff. In the following, we assume $n_a = (n,0)$ without loss of generality. The momentum integration and subsequent energy minimization is performed numerically. The resulting phase diagram assuming $\phi, n \ll \Lambda^2$ is shown in Fig.~\ref{fig:MFTphasediag}. 
If the interaction is predominantly $g$ ($g'$), the antiferromagnetic (nematic) state is preferred. While the electronic spectrum in the antiferromagnetic phase is fully gapped, in the nematic phase each quadratic band touching point splits into two gapless mini-Dirac cones. In between, however, a state in which both $\langle\phi\rangle$ and $\langle n \rangle$ are nonvanishing is stabilized---a coexistence phase. This phase is characterized by an anisotropic but fully gapped electronic spectrum; see inset in Fig.~\ref{fig:MFTphasediag}.

In the limit $g, g' \to 0$, the coexistence phase shrinks and is located around the line described by $g'=2g$ (dotted gray line in Fig.~\ref{fig:MFTphasediag}).
The latter can be understood from an RG perspective: The pertinent $\beta$ functions at one-loop order essentially%
\footnote{Note that in the strict mean-field limit $\Nf = \infty$, all that matters are the (anti-)commutation properties of the matrices appearing in the four-fermion interaction with the free fermion propagator. Hence, the fact that we are actually studying the spin counterpart of the rotationally invariant gapped state considered in Ref.~\cite{vafek10a} does not change the $\beta$ functions in this limit.}
follow from the $\Nf \to \infty$ limit of Eqs.~(7) and (8) of Ref.~\cite{vafek10a}. They are given by
\begin{align} \label{eq:flow-QBT}
	\beta_{g} = 2g^2, \qquad \beta_{g'} = (g')^2,
\end{align}
where we have rescaled $(g,g')\Nf/(4\pi) \mapsto (g,g')$ and dropped any terms that vanish for $\Nf \to \infty$. The $g$ axis, the $g'$ axis, and the line $g' = 2g$ are invariant subspaces of the RG flow.%
\footnote{The RG invariance of these subspaces may be understood by recognizing that they correspond to rays joining the Gaussian fixed point with certain (bi-)critical fixed points in $2+\varepsilon$ spatial dimensions, see Ref.~\cite{ray20}.}
We use conventions in which a positive $\beta$ function corresponds to an infrared relevant direction.
Hence, for positive initial couplings, the flow always diverges as $(g,g') \to (\infty,\infty)$ in the infrared. In fact, this occurs within finite RG time and signifies spontaneous symmetry breaking.
In the RG approach, the usual strategy to determine the nature of the symmetry-breaking ground state is based on comparing susceptibilities of the corresponding order parameters~\cite{vafek10a, vafek10b, cvetkovic12, lemonik12, janssen17a, boettcher17}.
We emphasize that such an analysis does not allow one to identify possible coexistence phases in a controlled way~\cite{cvetkovic12}. 
For the present large-$\Nf$ flow equations~\eqref{eq:flow-QBT}, the susceptibility analysis becomes particularly simple, as the ratio $g'/g$ approaches either zero or infinity in the infrared, depending on the initial values of the couplings: For $g'/g > 2$, we find that $g'/g \to \infty$ and the nematic susceptibility has the strongest divergence. For $g'/g < 2$, on the other hand, $g'/g \to 0$ and the antiferromagnetic susceptibility dominates. The RG invariant line $g'/g = 2$ hence represents the transition line between nematic and antiferromagnetic orders in the weakly interacting limit, in agreement with our mean-field analysis (see Fig.~\ref{fig:MFTphasediag}).
For finite short-range couplings, however, the mean-field calculation shows that the transition line is ``smeared out'' into an extended coexistence phase. Upon increasing $g$ and $g'$, the higher-order corrections incorporated in this calculation shift the location of the coexistence phase towards smaller ratio $g'/g$.
	
\begin{figure}
	\includegraphics[width=\linewidth]{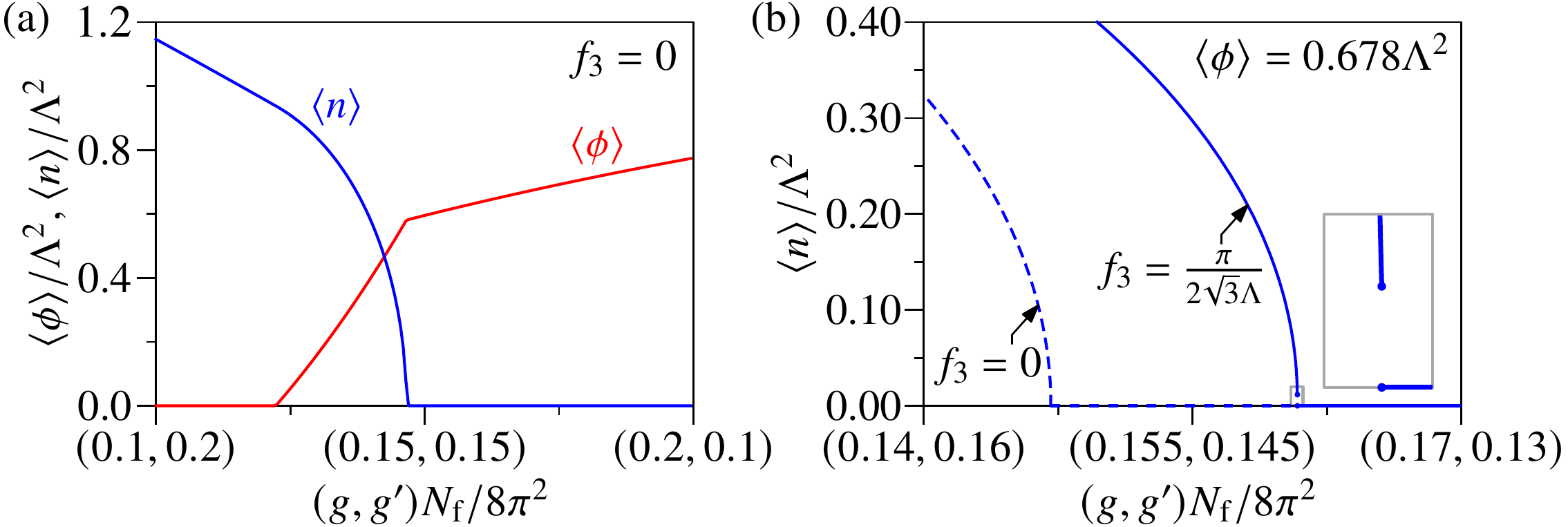}
	\caption{%
	(a) Nematic order parameter $\langle n \rangle$ (blue) and antiferromagnetic order parameter $\langle \phi \rangle$ (red) along the cut through parameter space indicated by the dashed black line in Fig.~\ref{fig:MFTphasediag}, in the mean-field approximation. In the model with continuous spatial rotational symmetry, both transitions into and out of the coexistence phase are continuous.
	(b) Nematic order parameter in the vicinity of the antiferromagnet-to-coexistence transition, showing the effects of the $f_3$ term defined in Eq.~\eqref{eq:f3}, which breaks the continuous spatial rotational symmetry down to 120$^\circ$ rotations on the honeycomb bilayer. Here, the antiferromagnetic order parameter $\langle \phi \rangle = 0.678\Lambda^2$ has been held constant for simplicity. The inset shows a zoom into the region very close to the transition (gray rectangle), illustrating the fact that finite $f_3 \neq 0$ renders the transition weakly first order.}
	\label{fig:order-parameters}
\end{figure}
	
Figure~\ref{fig:order-parameters} shows the evolution of the nematic and antiferromagnetic order parameters along the cut indicated by the dashed black line in Fig.~\ref{fig:MFTphasediag}.
In our simple model with continuous spatial rotational symmetry, both transitions into and out of the coexistence phase are continuous, see Fig.~\ref{fig:order-parameters}(a). In the vicinity of the nematic-to-coexistence transition, the antiferromagnetic order parameter develops an expectation value as $\langle \phi \rangle \propto g - g_\mathrm{c1}$ for $g \geq g_\mathrm{c1}$, where $g_\mathrm{c1}$ denotes the critical coupling. The linear behavior is consistent with Gross-Neveu-type quantum criticality in the large-$\Nf$ limit~\cite{hands93}.
Near the antiferromagnetic-to-coexistence transition at $g_\mathrm{c2} > g_\mathrm{c1}$, across which the fermionic spectrum retains a finite gap, the corresponding nematic order parameter has a square-root behavior, $\langle n \rangle \propto \sqrt{g_\mathrm{c2} - g}$ for $g \leq g_\mathrm{c2}$.
While this is consistent with the mean-field expectation for a purely bosonic transition, it turns out to be an artifact of our simple modeling, which assumes a continuous rotational symmetry. In fact, as shown in Fig.~\ref{fig:order-parameters}(b) and discussed in detail in Sec.~\ref{sec:afm-coexistence}, the antiferromagnetic-to-coexistence transition becomes weakly first order when perturbations, which break the continuous spatial rotational symmetry down to 120$^\circ$ rotations on the honeycomb bilayer, are taken into account.
The nematic-to-coexistence transition, by contrast, is expected to remain continuous upon the inclusion of such perturbations (see Sec.~\ref{sec:nem-coexistence}).

\begin{figure}
    \centering
	\includegraphics[scale=1.5]{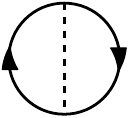}
	\caption{Simplest order-parameter fluctuation correction to the effective potential. The dashed (solid) lines refer to boson (fermion) propagators.}
	\label{fig:vacuumdiags}
\end{figure}

The present mean-field analysis represents the leading order of a systematic $1/\Nf$ expansion.
To incorporate the effects of order-parameter fluctuations on the effective potential at finite $\Nf$, one would need to evaluate higher-loop vacuum diagrams, the simplest of which is shown in Fig.~\ref{fig:vacuumdiags}. 
Such an analysis was performed in Ref.~\cite{ray20} for a single-order-parameter effective potential.
Alternatively, order-parameter fluctuations could be incorporated within a RG analysis along the lines of the works on competing orders in the monolayer case~\cite{classen15, classen16}.
Fluctuation effects may shift the phase boundaries in parameter space, but are expected to not alter our main conclusions concerning the existence of the coexistence phase and the order of the transitions into and out of this phase.
They do, however, play an important role for the critical behavior at the nematic-to-coexistence quantum critical point.
Instead of a comprehensive analysis of the full phase diagram, in this work we therefore restrict our study of the effects of order-parameter fluctuations to the vicinity of the nematic-to-coexistence transition. This is discussed in Sec.~\ref{sec:nem-coexistence}.
We also note that we have assumed $\phi, n \ll \Lambda^2$ in the above calculation, so that the integral in Eq.~\eqref{eq:VeffMFmaster} is dominated by universal logarithms such as $\ln(\Lambda^4/\phi^2)$ and $\ln(\Lambda^4/n^2)$. For larger interactions, this assumption no longer holds and nonuniversal effects that are beyond our effective analysis may become important. This is left for future work.

\section{Antiferromagnet-to-coexistence transition}
\label{sec:afm-coexistence}

In the antiferromagnetic phase with $\phi \neq 0$ and $n = 0$, the fermion spectrum is completely gapped out,
\begin{align}
	\varepsilon^\pm_{\phi,0}(\vec{p}) = \pm\sqrt{p^4 + \phi^2},
\end{align}
and the gap remains finite at the transition into the coexistence phase.
Consequently, the antiferromagnet-to-coexistence transition can be understood entirely within Ginzburg-Landau theory. The spontaneous breaking of spatial rotational symmetry is described by fluctuations of the two-component nematic order parameter $n_a$, $a=1,2$, which transforms as a second-rank tensor under spatial rotations, and acquires a vacuum expectation value across the transition.
However, it is important to note that the actual point group on the honeycomb bilayer includes only discrete $C_3$ rotations by 120$^\circ$ around a lattice site.
On general grounds, therefore, we should expect the effective potential to also include terms that are invariant only under the smaller discrete symmetry group $C_3$. The simplest such term is given by the cubic invariant $(n_1 + \rmi n_2)^3 + (n_1 - \rmi n_2)^3$, which is power-counting relevant compared to the quartic self-coupling.

While this term would ultimately be generated under RG flow, we can in fact show more: By including explicit rotational symmetry breaking at the level of the fermionic spectrum, we can derive the cubic invariant to appear in the effective potential explicitly. To this end, let us amend the fermionic Lagrangian~\eqref{eq:phenoL} by the irrelevant (in the usual power-counting sense) term
\begin{align} \label{eq:f3}
	\mathcal{L}_\text{QBT} \to \mathcal{L}_\text{QBT} + f_3 {\Psi}^\dagger \nabla^2 \rmi\bar{\partial}_a (\bar\Gamma_a \otimes \mathds{1}_2) \Psi,
\end{align}
where $(\bar{\partial}_a) = (\partial_x, -\partial_y)$, $a=1,2$, and $\bar{\Gamma}_a = (\tau_z \otimes \mathds{1}_2)\Gamma_a$.
This term follows naturally from the next-to-leading order expansion of the tight-binding dispersion near the $\vec K$ points in the Brillouin zone~\cite{pujari16}.
Identifying for simplicity the ultraviolet momentum cutoff $\Lambda$ with the inverse of the lattice constant $a_0$ as $\Lambda \sim \pi/a_0$, we obtain $f_3 \sim \pi/(2\sqrt{3} \Lambda)$ in our units~\cite{ray18}.
The term parametrized by $f_3$ is $C_3$ invariant, but not symmetric under the continuous rotation group in two spatial dimensions.
Let us now consider the mean-field effective potential for small $n_a \ll \phi$ in a finite antiferromagnetic background $\phi \neq 0$ and for finite $f_3 \neq 0$. Since the fermions are already gapped out, we run into no infrared divergences when Taylor expanding in $n_a$. Using polar coordinates $(n_a) = (n \cos2\vartheta,n\sin2\vartheta)$, we can write the effective potential for $n_a$ at the mean-field level in the form
\begin{multline}
	V_{\textnormal{MF}}^{(\phi)}(n,\vartheta) 
	= \frac{n^2}{2g'} 
	+ \frac{\Nf}{16\pi} \sum_{k,l = 0}^\infty \big[C_{kl,+}(\phi;f_3) \cos (2k\vartheta) \\ 
	+ C_{kl,-}(\phi;f_3) \sin (2k\vartheta)\big] n^{l+2},
	\label{eq:partialVefffull}
\end{multline}
with coefficients $C_{kl,\pm}$ that only depend on $\phi$ and $f_3$.
In Eq.~\eqref{eq:partialVefffull}, we have subtracted all $n$-independent offsets compared to Eq.~\eqref{eq:VeffMFmaster}, which are irrelevant for the present analysis. Following standard practice in Landau theory, we keep all terms up to and including $\mathcal{O}(n^4)$ (i.e., the lowest nontrivial order) in the effective potential.
The coefficients $C_{kl,\pm}(\phi;f_3)$ can be expanded in $f_3$, which allows us to evaluate all momentum integrals analytically to leading nontrivial order in $f_3$. This way, we finally arrive at the explicit result
\begin{align}
	V_{\textnormal{MF}}^{(\phi)}(n,\vartheta) 
	& \simeq 
	\frac{\Nf}{16\pi} \bigg\{
	\bigg[\frac{8\pi}{\Nf g'} + \frac14\left(\ln \frac14 \phi^2 - 2\right) + \frac{1}{8}f_3^2\bigg]n^2
	\nonumber\\&\phantom{{}={}} 
	+ f_3^2\left(\frac{1}{24} + \frac{1}{64}\ln \frac{1}{4}\phi^2\right) n^3 \cos(6\vartheta)
	\nonumber\\&\phantom{{}={}}
	+ \left(\frac{3}{32\phi^2} + \frac{1}{64} + \frac{9}{128}f_3^2\right)n^4\bigg\},
	\label{eq:Veffnofphi}
\end{align}
where we have rescaled $V_\text{MF}^{(\phi)}/\Lambda^4 \mapsto V_\text{MF}^{(\phi)}$, $\phi^2/\Lambda^4 \mapsto \phi^2$, $n^2/\Lambda^4 \mapsto n^2$, $f_3\Lambda \mapsto f_3$, and have kept only the leading- and subleading-order terms assuming the hierarchy $n \ll \phi \ll 1$. The latter assumption is consistent with small to intermediate $g$, since $\langle \phi \rangle \simeq \rme^{-2\pi/(g\Nf)}$ in mean-field theory \cite{sun09, ray18, ray20}.
We note that higher orders in $f_3^2$ also come with higher powers of $\phi$ for dimensional reasons; this defines \emph{a posteriori} the regime in which the expansion in $f_3$ is justified purely on grounds of its canonical dimension and independently of its value at the ultraviolet scale.
To be more precise, insertions of the $f_3$ term into the one-loop fermion bubble renders the integral increasingly ultraviolet divergent and infrared convergent; since finite $\phi$ is precisely what cures infrared divergences in this theory, the faster a given loop integrand vanishes in the limit of vanishing loop momenta, the faster its integral vanishes for $\phi \to 0$.

The middle term in Eq.~\eqref{eq:Veffnofphi} manifests the explicit symmetry breaking at the level of the effective potential for $n_a$: $\cos(6\vartheta)$ is only invariant under $\vartheta \mapsto \vartheta + \pi k/3$, $k\in \mathds{Z}$. Minimizing the potential with respect to $\vartheta$, we find that the orientation of the vector $n_a$ locks on to $\vartheta = 0$ at the minimum for sufficiently small $\phi$, while $\vartheta = \pi/6$ for larger values of $\phi$.
Importantly, the presence of the cubic term renders the antiferromagnetic-to-coexistence transition first order, with the jump discontinuity at the phase boundary working out to
\begin{align}
	\delta \langle n \rangle = \frac{1}{12}f_3^2 \langle\phi\rangle^2 \left(\ln\frac{1}{4\langle\phi\rangle^2}-\frac{8}{3}\right) + O(\langle\phi\rangle^4 \ln \langle \phi \rangle^2).
\end{align}
Note that the above implies the emergence of a hierarchy among the order parameters, $\delta\langle n \rangle \sim \langle\phi\rangle^2 \ln \langle\phi\rangle^{-2} \ll \langle \phi \rangle$ in the limit of small interaction strengths. Technically, the smallness of $\delta \langle n \rangle$ justifies expanding its effective potential in powers of $n$, as well as treating $\phi$ as a background field with no backreaction from $n$, even though the transition itself is not continuous. Physically, $\delta\langle n\rangle$ measures how badly the transition fails to be continuous. Since $\delta n \to 0$ for $g \to 0$, the transition is only weakly first order for small to intermediate four-fermion coupling, as also illustrated numerically in Fig.~\ref{fig:order-parameters}(b).

\section{Nematic-to-coexistence transition}
\label{sec:nem-coexistence}

In the nematic phase, the low-energy excitations are massless Dirac fermions. They acquire a full gap in the coexistence phase.
At the transition, the antiferromagnetic order parameter $\vec\phi$ becomes critical.
The Gross-Neveu-Heisenberg (= chiral Heisenberg) universality class~\cite{rosenstein93, herbut09b, janssen14, zerf17, knorr18, gracey18} is therefore a natural candidate to describe the continuous nematic-to-coexistence transition.
The purpose of this section is to confirm this scenario.
This Gross-Neveu-Heisenberg universality class is characterized by a dynamical critical exponent $z=1$ and a relativistic symmetry in (2+1)-dimensional spacetime.
In the noninteracting limit, our model has $z=2$, reflecting the nonrelativistic dispersion in the Luttinger semimetal state~\cite{ray20}. As a consequence of the finite background nematic order present across the nematic-to-coexistence transition, the system does not even feature explicit discrete $C_3$ rotational symmetry near the transition.
Nevertheless, in this section, we demonstrate that not only a continuous rotational symmetry, but even a full relativistic symmetry in 2+1 space-time dimensions becomes emergent at the quantum critical point at low energy.
To this end, we systematically study the fate of perturbations that break both rotational and space-time symmetries of the relativistic subspace of theory space.
In particular, we show that such symmetry-breaking perturbations are always RG irrelevant near the Gross-Neveu-Heisenberg fixed point.
The nematic-to-coexistence transition on the honeycomb bilayer therefore falls into the same family of phase transitions known from the monolayer system. There is, nevertheless, one important difference: As each quadratic band touching point on the bilayer splits into two mini-Dirac cones in the nematic phase, the number of fermion flavors is doubled in comparison to the semimetal-to-antiferromagnetic transition on the monolayer.
In the physical situation for spin-$1/2$ fermions on the Bernal-stacked honeycomb bilayer, we have $\Nf=4$ flavors of two-component Luttinger fermions in the noninteracting limit. In the nematic phase, this then leads to $\Nf = 4$ flavors of \emph{four}-component Dirac fermions (in other words, the total number of Dirac spinor components is $4\Nf = 16$).

This section is divided into three parts: In Sec.~\ref{subsec:4Fermi}, we demonstrate the emergence of relativistic symmetry within an $\varepsilon$ expansion around the lower critical space-time dimension of $D_\text{low}=2$.
Section~\ref{subsec:GNY} contains an analysis using an $\epsilon$ expansion around the upper critical space-time dimension of $D_\text{up}=4$, with a consistent result.
These two approaches complement each other, as the least-irrelevant symmetry-breaking perturbations are of different natures near the lower and upper critical dimensions.
The resulting Gross-Neveu-Heisenberg critical exponents relevant for the present situation are discussed in Sec.~\ref{subsec:exponents}.

\subsection{Emergent Lorentz symmetry: Expansion near lower critical dimension}
\label{subsec:4Fermi}

Unlike the Gross-Neveu-Ising case, in the Gross-Neveu-Heisenberg universality class, the renormalization of the pertinent four-fermion interaction is not closed. Already at the one-loop level, a spin current interaction is generated, which in turn generates further interactions. The upshot is that the renormalization of the chiral Heisenberg universality class in the vicinity of the lower critical space-time dimension $D_\text{low} = 2$ has so far not been systematically studied, even at one loop. Before investigating its stability with respect to perturbations, we hence need to first identify the Gross-Neveu-Heisenberg fixed point in the multidimensional space of four-fermion couplings.

\subsubsection{Minimal four-fermion model}
As a first step, we aim at establishing a basis in the space of four-fermion couplings. To this end, we classify all possible four-fermion interactions in terms of symmetry.
%
%
In order to retain the spinor structure relevant to the physical situation in bilayer graphene, we devise the minimal four-fermion model in fixed $D = 2+1$ space-time dimensions. The dimensional continuation to noninteger dimensions will be discussed in the context of the loop integration in Sec.~\ref{par:2+eps-GNY}.
In accordance with previous works~\cite{herbut09a, vafek10a}, we employ a four-dimensional reducible representation of the Clifford algebra $\{\gamma_\mu , \gamma_\nu\} = 2\delta_{\mu\nu}\mathds{1}_4$ with $\mu,\nu = 0,1,2$.
In addition, there exist two anticommuting matrices $\{\gamma_3,\gamma_\mu\} = \{\gamma_5,\gamma_\mu\} = \{\gamma_3,\gamma_5\} = 0$. Finally, a customary shorthand is $\gamma_{35} \coloneq \rmi \gamma_3 \gamma_5$.

The Gross-Neveu-Heisenberg four-fermion interaction can then be written as $[\overline{\psi}_a(\mathds{1}_{4} \otimes \sigma_\alpha)\psi_a]^2$, where $\mathds{1}_4$ acts on the layer and $\vec K$-point degrees of freedom and the Pauli matrices $\sigma_\alpha$, $\alpha = 1,2,3$, act on the spin degree of freedom of the eight-component spinor $\psi_a$. Furthermore, the flavor index $a=1,2$ corresponds to the two mini-Dirac cones that develop at both $\vec K$ points due to the background nematic order.
For simplicity, we restrict ourselves to interactions that have a singlet structure with respect to the flavor index $a$. 
There are then \emph{a priori} 64 independent four-fermion interactions,
\begin{align}
	\sum_{M \in \mathcal{B}} G_M (\overline{\psi}_a M \psi_a)^2,
	\end{align}
where $M$ are complex $8 \times 8$ matrices and $\mathcal{B} = \mathcal{B}_\text{s} \cup \mathcal{B}_\text{v}$ is a basis of $\mathds{C}^{8\times 8}$, with
$\mathcal{B}_\text{s} = \{ \mathds{1}_4, \gamma_\mu, \gamma_3, \gamma_5, \gamma_\mu \gamma_3, \gamma_\mu \gamma_5, \gamma_{35}, \gamma_\mu\gamma_{35}\}  \otimes \mathds{1}_2$
corresponding to the scalar spin sector and 
$\mathcal{B}_\text{v} = \{ \mathds{1}_4, \gamma_\mu, \gamma_3, \gamma_5, \gamma_\mu \gamma_3, \gamma_\mu \gamma_5, \gamma_{35}, \gamma_\mu\gamma_{35}\}  \otimes \sigma_\alpha$
to the vector spin sector, with $\mu = 0,1,2$ and $\alpha = x,y,z$.
Following the procedure outlined in Ref.~\cite{herbut09a}, the number of independent interactions may be whittled down systematically as follows:
\paragraph{Lorentz and $\SUtwo$ spin symmetry} The four-fermion interactions may be grouped according to their behavior under Lorentz and $\SUtwo$ spin transformations. To be precise, if the subset $\mathcal{A} \subset \mathcal{B}$ is invariant under combined Lorentz and $\SUtwo$ transformations, then $G_{K} = G_{L}$ for all $K,L \in \mathcal{A}$. The decomposition of $\mathcal{B}$ into disjoint subsets $\mathcal{A}$ is almost entirely taken care of automatically above by grouping them according to Lorentz and $\SUtwo$ indices. The grouping is almost exhaustive: For a final symmetry reduction, we need to take into account that $(\gamma_3,\gamma_5)$ is a vector under the $\text{U}(1)$ chiral symmetry generated by $\gamma_{35}$, which corresponds to translational invariance~\cite{herbut09a}. There are hence at most twelve symmetry-independent couplings.

\paragraph{Fierz identities} To further reduce the number of couplings, we need to exploit Fierz identities~\cite{herbut09a, gies10, gehring15}. Given an orthonormal basis $\operatorname{Tr}(MN) = \delta_{MN}\operatorname{Tr}\mathds{1}$, we have for Grassmann fields
\begin{align} \label{eq:Fierz}
	(\overline{\psi}_a M\psi_a)^2 = \sum_{N\in\mathcal{B}} \frac{-1}{64}\operatorname{Tr}(MNMN)(\overline{\psi}_a N\psi_a)^2.
\end{align}
Gathering the symmetry-independent interactions into a twelve-dimensional vector $\vec u$, the above may be recast into the form $F \vec u = 0$ with with a $12\times 12$ Fierz matrix $F$. It turns out that the Fierz matrix $F$ thus constructed has six zero eigenvalues; there are hence only six independent couplings after symmetry and Fierz reduction. We thus arrive at
\begin{align} \label{eq:L-GNH}
	\mathcal{L}_{\text{GNH}} &= 
	\overline{\psi}_a (\gamma_\mu \otimes \mathds{1}_2) \partial_\mu \psi_a 
	\nonumber\\&\phantom{{}={}} {} 
	-\frac{G_1}{2 \Nf} [\overline{\psi}_a (\mathds{1}_4 \otimes \sigma_\alpha)\psi_a]^2
	-\frac{G_2}{2 \Nf} [\overline{\psi}_a(\gamma_{\mu} \otimes \sigma_\alpha) \psi_a]^2 
	\nonumber\\&\phantom{{}={}} {}
	-\frac{G_3}{2 \Nf} [\overline{\psi}_a(\gamma_{35} \otimes \sigma_\alpha) \psi_a]^2
	-\frac{G_4}{2 \Nf}(\overline{\psi}_a \psi_a)^2 
	\nonumber\\&\phantom{{}={}} {}
	- \frac{G_5}{2 \Nf} [\overline{\psi}_a(\gamma_{\mu}\otimes \mathds{1}_2) \psi_a]^2 
	- \frac{G_6}{2 \Nf} [\overline{\psi}_a(\gamma_{35} \otimes \mathds{1}_2) \psi_a]^2
\end{align}
as a minimal four-fermion theory in which to embed the Gross-Neveu-Heisenberg fixed point. Note that we have reinstated the flavor number $\Nf$, corresponding to the number of four-component Dirac spinors, with $a=1,\dots,\Nf/2$ for $\Nf$ even. We reiterate that the case relevant for the nematic-to-coexistence transition of spin-$1/2$ fermions on the honeycomb bilayer corresponds to $\Nf = 4$.

\subsubsection{Gross-Neveu-Heisenberg fixed point}
\label{par:2+eps-GNY}
To obtain the RG flow of the couplings $G_1, \dots, G_6$ in Eq.~\eqref{eq:L-GNH}, we have to perform the loop integration. Here, we evaluate the angular integrals in fixed $D=2+1$ space-time dimensions, while the dimensions of the couplings are counted in general dimension~\cite{vojta00, janssen17a}. This allows us to retain the spinor structure of the physical system in $d=2$ spatial dimensions.
We have obtained the flow equations at one-loop order by applying the general formula given in Ref.~\cite{gehring15}. In addition, we have performed a large-$\Nf$ expansion of the one-loop flow equations, for three reasons: (i) tractability, in that solutions of fixed-point equations can be found analytically in its entirety, with human-readable results; (ii) transparency, in that relations to mean-field theory ($\Nf \to \infty$) become more readily apparent; and (iii) simplicity, in that the fixed point pertaining to the $\SUtwo$-symmetry-breaking transition is unambiguously identifiable. Let us expand on this last point a little: At general $\Nf$, a fixed point generically has many nonzero four-fermion couplings. Determining unambiguously which one among the many fixed points pertains to the Gross-Neveu-Heisenberg universality class for arbitrary $\Nf$ is typically a laborious exercise, entailing the computation of scaling dimensions of every conceivable bilinear at every fixed point. In the large-$\Nf$ limit, however, this is unambiguous (and essentially known already from mean-field theory): the Gross-Neveu-Heisenberg universality class is governed by the fixed point satisfying $G_1 = \mathcal{O}(1)$ and $G_{i\neq 1} = \mathcal{O}(1/\Nf)$. We expect this large-$\Nf$ argument to be sufficient for the present case $\Nf = 4$ and leave the full investigation for arbitrary $\Nf$ to future work.

For the Gross-Neveu-Heisenberg fixed point in $D=2+\varepsilon$ dimensions, we obtain the fixed-point couplings
\begin{align}
	G_{1,\star} &= \left(\frac{2}{3} - \frac{2}{3\Nf} + \frac{2}{9\Nf} + \frac{274}{81\Nf^3}\right)\varepsilon + \mathcal{O}\left(\varepsilon^2,1/\Nf^4\right), \displaybreak[0]\\
	G_{2,\star} &= \left(-\frac{4}{9\Nf} + \frac{76}{81 \Nf^2} - \frac{860}{729\Nf^3}\right)\varepsilon + \mathcal{O}\left(\varepsilon^2,1/\Nf^4\right), \displaybreak[0] \\
	G_{3,\star} &= -\frac{40 \varepsilon}{9\Nf^3} + \mathcal{O}\left(\varepsilon^2,1/\Nf^4\right), \displaybreak[0] \\
	G_{4,\star} &= -\frac{8\varepsilon }{3\Nf^3} + \mathcal{O}\left(\varepsilon^2,1/\Nf^4\right), \displaybreak[0] \\
	G_{5,\star} &= \frac{8 \varepsilon }{9\Nf^3} + \mathcal{O}\left(\varepsilon^2,1/\Nf^4\right), \displaybreak[0] \\
	G_{6,\star} &= -\frac{8 \varepsilon }{3\Nf^3} + \mathcal{O}\left(\varepsilon^2,1/\Nf^4\right).
\end{align}
We have explicitly checked that this fixed point indeed features precisely one infrared relevant direction in the six-dimensional theory space parametrized by $G_1, \dots, G_6$, hence corresponding to a quantum critical point.
Note that starting with the Heisenberg channel, a second channel is immediately generated at first subleading order, $O(1/\Nf)$. This is the four-fermion interaction $[\overline{\psi}_a(\gamma_{\mu} \otimes \sigma_\alpha) \psi_a]^2$, the $\SUtwo$-vector counterpart of the conventional [$\SUtwo$-scalar] Thirring interaction. From $\mathcal{O}(1/\Nf^3)$ onwards, all channels get involved.

\subsubsection{Fate of rotational symmetry breaking}
We are now in the position to study the fate of rotational anisotropies under the RG flow. 
Since the background nematic order respects inversion and lattice translational symmetries, it is sufficient to restrict the discussion to perturbations that leave discrete symmetries intact and break explicitly only the continuous rotational symmetry of the Gross-Neveu-Heisenberg fixed point.
Rotational-symmetry-breaking terms in the quadratic part of the fermionic Lagrangian, such as anisotropic Fermi velocities, are marginal within the one-loop expansion considered here. Their relevance (or lack thereof) will be studied within the Gross-Neveu-Yukawa-Heisenberg model discussed in Sec.~\ref{subsec:GNY}.
Here, we focus on perturbations in the interacting quartic part of the Lagrangian. Restricting ourselves to small deviations away from the Gross-Neveu-Heisenberg fixed point, we may use the Fierz identities of Eq.~\eqref{eq:Fierz} to write an arbitrary rotational-symmetry-breaking four-fermion interaction as a linear combination of four basis terms parametrized as
\begin{align}
	\mathcal{L}_{\textnormal{GNH}}' & =  \mathcal{L}_{{\textnormal{GNH}}} 
	\nonumber \\&\phantom{{}={}} {}
	- \tfrac12\delta_1 \left\{[\overline{\psi}_a(\gamma_1\otimes \mathds{1}_2) \psi_a]^2 - [\overline{\psi}_a(\gamma_0\otimes \mathds{1}_2) \psi_a]^2\right\}
	\nonumber \\&\phantom{{}={}} {}
	- \tfrac12\delta_2 \left\{[\overline{\psi}_a(\gamma_2\otimes \mathds{1}_2) \psi_a]^2 - [\overline{\psi}_a(\gamma_0\otimes \mathds{1}_2) \psi_a]^2\right\}
	\nonumber \\ &\phantom{{}={}} {}
	- \tfrac12\delta_3 \left\{[\overline{\psi}_a(\gamma_1\otimes\sigma_\alpha) \psi_a]^2 - [\overline{\psi}_a(\gamma_0\otimes\sigma_\alpha) \psi_a\right]^2\} \nonumber \\ &\phantom{{}={}} {}
	- \tfrac12\delta_4 \left\{[\overline{\psi}_a(\gamma_2\otimes\sigma_\alpha) \psi_a]^2 - [\overline{\psi}_a(\gamma_0\otimes\sigma_\alpha) \psi_a]^2\right\},
\end{align}
with couplings $\delta_1, \dots, \delta_4$.
Using again the general formula of Ref.~\cite{gehring15}, we find the four eigenvalues of the stability matrix $(\partial \beta_{\delta_i}/\partial \delta_j)$ at the Gross-Neveu-Heisenberg fixed point as
\begin{align}
    \theta_1 &= \left(-1 -\frac{1}{\Nf}+\frac{7}{3 \Nf^2}+\frac{25}{54 \Nf^3}\right)\varepsilon + \mathcal{O}(\varepsilon^2, 1/\Nf^4),\\
    \theta_2 &= \left(-1 + \frac{1}{\Nf}-\frac{7}{3 \Nf^2}-\frac{7}{54 \Nf^3}\right)\varepsilon
	+ \mathcal{O}(\varepsilon^2, 1/\Nf^4),\\
    \theta_3 &= \left(-1 -\frac{1}{9 \Nf}+\frac{121}{81 \Nf^2}-\frac{3943}{1458 \Nf^3}\right)\varepsilon + \mathcal{O}(\varepsilon^2, 1/\Nf^4),\\
    \theta_4 &= \left(-1 + \frac{1}{9\Nf}-\frac{157}{81 \Nf^2}+\frac{4825}{1458 \Nf^3}\right)\varepsilon + \mathcal{O}(\varepsilon^2, 1/\Nf^4).
\end{align}
For $\Nf = 4$, pertaining to the present case of nematic-to-coexistence transition on the honeycomb bilayer, all four eigenvalues are negative. Hence, at the transition, not just rotational symmetry, but also Lorentz symmetry is emergent in the infrared.

\subsection{Emergent Lorentz symmetry: Expansion near upper critical dimension}
\label{subsec:GNY}

The above one-loop four-fermion results are \emph{a priori} valid only in the vicinity of the lower critical space-time dimension of two. In $2+1$ space-time dimensions, corrections from higher loop orders may be sizeable. To check the robustness of our conclusions, we now consider rotational-symmetry-breaking perturbations in the opposite limit near the upper critical space-time dimension $D_\text{up} = 4$.

\subsubsection{Gross-Neveu-Yukawa-Heisenberg model}
The renormalizable field theory in this limit is the Gross-Neveu-Yukawa-Heisenberg model with Lagrangian
\begin{align} \label{eq:L-GNYH}
	\mathcal{L}_{\text{GNYH}} &= 
	\overline{\psi}_a [\mathds{1}_4 \otimes (\gamma_0\partial_0 + v_x \gamma_1 \partial_1 + v_y \gamma_2\partial_2)]\psi_a
	+ \tfrac12 (\partial_\mu\phi_\alpha)^2
	\nonumber \\ &\quad
	- h \phi_\alpha \overline{\psi}_a \left(\mathds{1}_{4} \otimes \sigma_\alpha\right)\psi_a
	+ \lambda (\phi_\alpha \phi_\alpha)^2,
\end{align}
where $a=1,\dots,\Nf/2$, in agreement with the representation used in Sec.~\ref{subsec:4Fermi}.
In Eq.~\eqref{eq:L-GNYH}, spatial rotational symmetry breaking is encoded in the direction-dependent Fermi velocities $v_x$ and $v_y$. Their bosonic counterparts $c_x$ and $c_y$ can be subsumed into direction-dependent dynamical critical exponents, see below, and we have hence set $c_x$ and $c_y$ to unity from the outset.
We have replaced the four-fermion interaction parametrized by $G_1$ in Sec.~\ref{subsec:4Fermi} by a Yukawa interaction between the fermions and the $\SUtwo$ order-parameter field $\phi_\alpha$, parametrized by the coupling $h$, which becomes marginal at the upper critical space-time dimension $D_\text{up} = 4$.
The bosonic self-interaction with coupling $\lambda$ is generated by the RG and has therefore been included as well. It also becomes marginal at the upper critical dimension.

In order to deal with the spatial anisotropy, we perform field-theoretic RG, with loop integrals carried out over all momenta. In the spirit of the $\epsilon$ expansion, we formally extend the \emph{time} axis to a $(2-\epsilon)$-dimensional Euclidean space, keeping the spatial dimension $d=2$ fixed and assuming that all integrands have been symmetrized in frequency $q_0$ before the dimensional continuation. The self-energy diagrams, which are the main subjects of study in this subsection, will turn out to be infrared divergent after expanding in powers of external momenta, which we regularize with a cutoff; the renormalization scale $\mu$ is introduced thus. The measure of the loop integration can thus be written as
\begin{align}
	\int_q \coloneq \int_{|q_0| > \mu}\frac{\rmd^{2-\epsilon} q_0\,\rmd^2 \vec{q}}{(2\pi)^{4-\epsilon}}.
\end{align}
The terms in the Lagrangian $\mathcal L_\text{GNYH}$ are accordingly promoted to bare quantities, with $v_i \to Z_{v_i} v_i$ ($i \in \{x,y\}$) and $\Phi \to \sqrt{Z_\Phi}\Phi$ ($\Phi \in \{ \phi, \psi \}$). We absorb the running of the bosonic velocities $c_i$ into ``inverse dynamical critical exponents,'' $\partial_i \to Z_{p,i} \partial_i$, where the $Z_{p,i}$ parametrize the relative scaling of momentum coordinates with respect to frequency. In other words, we measure the Fermi velocities $v_x$ and $v_y$ in units of $c_x$ and $c_y$, respectively.
Finally, since our regularization scheme breaks Lorentz invariance, we need nonmultiplicative counterterms, such as
\begin{align} \label{eq:counterterms}
	\mathcal{L}_{\text{rest.}} = D_\psi\,\overline{\psi} \partial_0 \gamma_0 \psi + \tfrac12 D_\phi\,\phi (-\partial_0^2)\phi.
\end{align}
These are required to ensure that in the Lorentz-invariant limit $v_x = v_y = 1$ there is no residual breaking of Lorentz symmetry (which would then solely be a regularization artefact).
This is reminiscent of the treatment of supersymmetric theories, where the often-used dimensional regularization breaks supersymmetry, and one has to resort to nonmultiplicative counterterms to restore it~\cite{hollik01}.
Just like usual multiplicative counterterms, these counterterms are often not unique, but can be judiciously constrained by demanding certain properties of the regularization procedure (see Appendix~\ref{app:selfenerg}).
Equation~\eqref{eq:counterterms} represents the simplest choice that is sufficient for our purposes.

\subsubsection{Gross-Neveu-Heisenberg fixed point}

The theory defined by Eq.~\eqref{eq:L-GNYH} features an interacting fixed point located within the relativistic subspace $v_{x,\star} = v_{y,\star} = 1$ at
\begin{align} \label{eq:GNY-fixedpoint}
    h_\star^2 & = \frac{\pi}{\Nf + 1}\epsilon + \mathcal O(\epsilon^2), \\
    \lambda_{\star} & = \frac{\pi}{22}\left(-1+\frac{2+\sqrt{1+\Nf (\Nf + 9)}}{\Nf + 1}\right)\epsilon + \mathcal O(\epsilon^2),
\end{align}
where $\epsilon = 4-D$ and we have rescaled $\mu^{-\epsilon} S_{2-\epsilon}(2\pi)^{\epsilon - 2} h^2 \mapsto h^2$ and $\mu^{-\epsilon} S_{2-\epsilon}(2\pi)^{\epsilon - 2} \lambda \mapsto \lambda$, with $S_{2-\epsilon}$ being the surface area of the unit sphere in $2-\epsilon$ dimensions.
We note that the above fixed-point values are regularization dependent and cannot be obtained by a simple rescaling of the corresponding values within, say, the Wilson scheme~\cite{janssen18}.
As is well known~\cite{herbut09b, janssen14, zerf17, janssen18}, in the vicinity of the Gross-Neveu-Heisenberg fixed point, the only relativistic-symmetry-allowed perturbation that is RG relevant towards the infrared is the quadratic term $\phi_\alpha \phi_\alpha$, which corresponds to the tuning parameter of the transition.
Within the critical hyperplane, in which this term is tuned to vanish in the renormalized action by definition, the Gross-Neveu-Heisenberg fixed point is hence stable.
In the following, we show that the stability holds also when small perturbations that break the rotational symmetry are taken into account. The fixed point hence features emergent Lorentz invariance in the low-energy limit.

\subsubsection{Fate of rotational symmetry breaking}

\begin{figure}[tb]
	\includegraphics[scale=1]{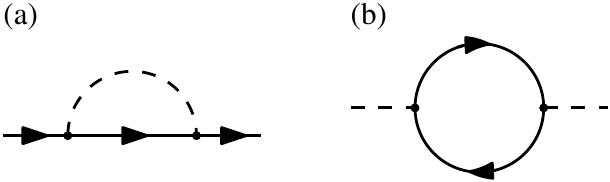}
	\caption{(a) Bosonic and (b) fermionic self-energy Feynman diagrams.}
	\label{fig:selfenerg}
\end{figure}

Within the Gross-Neveu-Yukawa-Heisenberg model, the fate of rotational symmetry breaking is determined by the flows of the Fermi velocities $v_x$ and $v_y$ in units of the boson velocities $c_x$ and $c_y$.
The corresponding self-energy diagrams at one-loop order are shown in Fig.~\ref{fig:selfenerg}. Details of the evaluation of these diagrams are deferred to Appendix~\ref{app:selfenerg}.
Defining $\beta_{v_i} \equiv -\mu \rmd v_i/\rmd \mu$ in terms of the RG scale $\mu$, we find
\begin{align}
	\beta_{v_x} &= \frac{h^2}{2\pi} 
	\biggl\{
	\frac{1-v_x^2}{v_y} \Nf
	\nonumber\\&\phantom{{}={}} {}
	+ \left[4\pi I_{211}(v_x, v_y) - 4\pi I_{210}(v_x,v_y) + \tfrac{1}{4}\right]v_x\Nb  
	\biggr\}, \label{eq:betavx} \displaybreak[1]\\
	\beta_{v_y} &= \frac{h^2}{2\pi}
	\biggl\{
	\frac{1-v_{y}^2}{v_{x}} \Nf 
	\nonumber\\&\phantom{{}={}} {}
	+ \left[4\pi I_{211}(v_{y}, v_{x}) - 4\pi I_{210}(v_{y},v_{x}) + \tfrac{1}{4}\right]v_{y}\Nb 
	\biggr\}, \label{eq:betavy}
\end{align}
where $I_{211}(v_x,v_y)$ and $I_{210}(v_x,v_y)$ are functions of the velocities $v_x$ and $v_y$ alone. They are defined in Appendix~\ref{app:selfenerg} and and explicit forms are given in Appendix~\ref{app:mastint}.
In Eqs.~\eqref{eq:betavx} and \eqref{eq:betavy}, we have employed the same rescaling of the Yukawa coupling as stated below Eq.~\eqref{eq:GNY-fixedpoint}.
As before, $\Nf$ counts the number of four-component Dirac fermions, with $\Nf=4$ for the case relevant for the nematic-to-coexistence transition on the honeycomb bilayer.
For generality, we have also introduced a generic number $\Nb$ of boson species, which allows one to easily adapt the current analysis to Gross-Neveu-Yukawa models with a different number of order-parameter components. For the antiferromagnetic order parameter discussed in this work, we have $\Nb = 3$.
Note that Eq.~\eqref{eq:betavy} can be obtained from Eq.~\eqref{eq:betavx} by exchanging $v_x \leftrightarrow v_y$ and vice versa.
 
The constraint $v_x \equiv v_y \eqqcolon v$ defines the rotationally symmetric subspace, which is invariant under RG flow for symmetry reasons. The $\beta$ function for the rotationally invariant Fermi velocity $v$ within this subspace reads explicitly
\begin{align}
	\beta_v &= \frac{h^2}{2\pi}\left[\frac{1-v^2}{v}\Nf  \right.\nonumber\\
	&\phantom{{}={}} \left. {} + \frac{v^4 +4v^2  - 5 - 2\left(1+2v^2\right)\ln v^2}{4(1-v^2)^2} v \Nb \right].
\end{align}
In the vicinity of the relativistic fixed point at $v_\star = 1$, the flow of the Fermi velocity can be expanded as
\begin{align}
    \beta_v = - \frac{\Nf h^2}{\pi} (v-1)  + \mathcal O((v-1)^2).
\end{align}
Within the rotationally invariant subspace, the relativistic Gross-Neveu-Heisenberg fixed point is therefore stable, in agreement with previous results for similar models~\cite{roy16, janssen17b}.

\begin{figure}[tb]
	\includegraphics[width=.8\columnwidth]{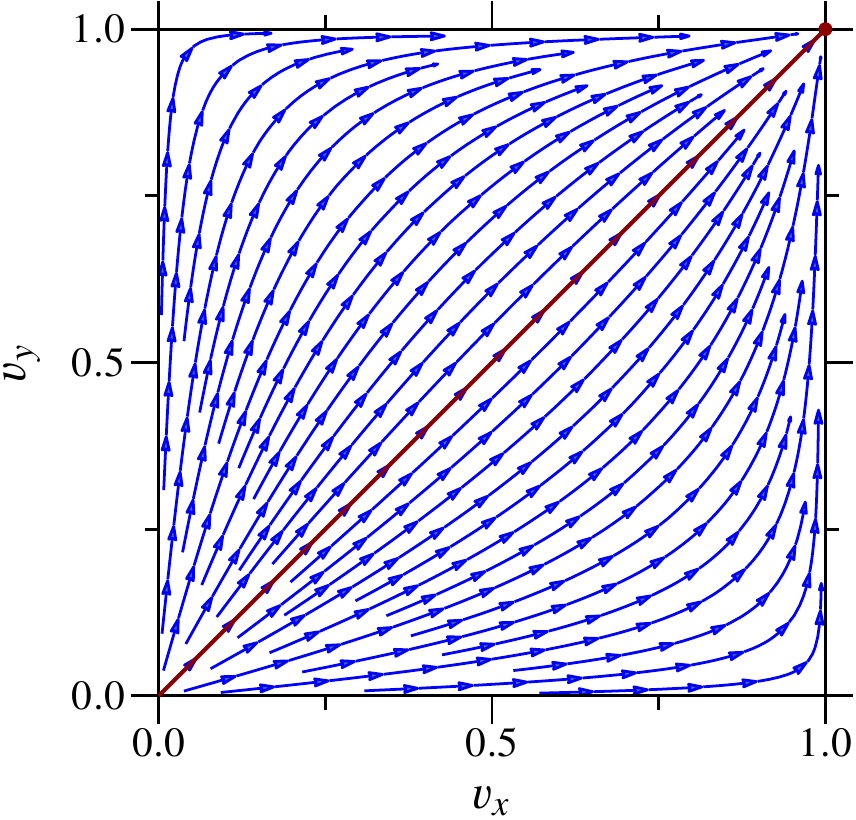}
	\caption{RG flow of Fermi velocities $v_x$ and $v_y$ in units of the boson velocities $c_x$ and $c_y$ for $\Nf = 4$ and $\Nb  = 3$ at one-loop order. Dark red line represents the rotationally-invariant subspace $v \coloneqq v_x = v_y$. All points flow ultimately to the relativistic fixed point $v_\star = 1$ (dark red point), even though flow lines initially ``fan out'' from the $v_x = v_y$ line for small enough initial values.}
	\label{fig:streamplt}
\end{figure}

To study whether the rotationally invariant subspace is stable or not with respect to rotational-symmetry-breaking perturbations, we set $v_x = v$ and $v_y = (1 + \delta)v$, and expand $\beta_\delta = \beta_{v_y} - \beta_{v_x}$ to first order in the anisotropy parameter $\delta$. For small $\delta \ll 1$, we thus obtain
\begin{align}
	\beta_\delta &= -\frac{h^2}{2\pi}\left[\vphantom{\left(\frac{2\left(1 + v^2 + 4 v^4\right) \ln v^2}{(1-v^2)^3}\right)}\left(3v - \frac{1}{v}\right)\Nf  \right.\nonumber\\&\phantom{{}={}} \left. {} 
	- \left(\frac{\left(1 + v^2 + 4 v^4\right) \ln v^2}{2(1-v^2)^3} + \frac{v^4+10 v^2+1}{4(1-v^2)^2}\right)\Nb \right]\delta
	\nonumber\\&\phantom{{}={}} {}
	+\mathcal O(\delta^2).
\end{align}
Near the relativistic fixed point at $v_\star = 1$, we find
\begin{align}
    \beta_\delta = -\frac{h^2}{4\pi}(4\Nf + \Nb)\delta+\mathcal O(\delta^2,v-1).
\end{align}
Importantly, a small rotational anisotropy is therefore irrelevant in the sense of the RG.
In the vicinity of the Gross-Neveu-Heisenberg fixed point, the relativistic symmetry hence remains emergent even when a rotational anisotropy is symmetry-allowed on the microscopic level.
This is illustrated in Fig.~\ref{fig:streamplt}, which shows the RG flow of $v_x$ and $v_y$ using the full $\beta$ functions of Eqs.~\eqref{eq:betavx} and \eqref{eq:betavy}.
All points flow ultimately to the relativistic fixed point $v_{x,\star} = v_{y,\star} = 1$, even though flow lines initially ``fan out'' from the rotationally-invariant subspace line for small enough initial values.
In agreement with the result from the $2+\varepsilon$ expansion discussed in the previous subsection, we conclude that the nematic-to-coexistence transition on the honeycomb bilayer features emergent Lorentz symmetry and is described by the relativistic Gross-Neveu-Heisenberg fixed point.
	
\subsection{Critical exponents}
\label{subsec:exponents}

We conclude this section by providing estimates for the critical exponents characterizing the nematic-to-coexistence transition on the honeycomb bilayer. 
As we have established above, the transition falls into the Gross-Neveu-Heisenberg universality class with $\Nf = 4$ flavors of four-component Dirac fermions.
As a consequence of the emergent Lorentz invariance, the transition is characterized by a unique dynamical critical exponent
\begin{align}
    z = 1.
\end{align}

Further, we discuss the correlation-length exponent $\nu$, the order-parameter anomalous dimension $\eta_\phi$, and the fermion anomalous dimension $\eta_\psi$.
Higher-order corrections in perturbation theory are available in the literature for this universality class to fourth order in the $4-\epsilon$ expansion~\cite{zerf17} and to second order in the large-$\Nf$ expansion, with the fermion anomalous dimension known up to third order~\cite{gracey18, gracey21}. 
Furthermore, a nonperturbative calculation using functional RG in the improved local potential approximation is also available \cite{janssen14}. Here, we perform the necessary post-processing of these previous results to provide combined theoretical estimates for the nematic-to-coexistence transition on the honeycomb bilayer.

First, let us consider the series expansions.
In fixed $D=2+1$ space-time dimensions, the large-$\Nf$ expansions of the exponents are~\cite{gracey18}
\begin{align}
    1/\nu & = 1 - \frac{8}{\pi^2 \Nf} + \frac{36\pi^2 + 416}{3\pi^4 \Nf^2} + \mathcal O(1/\Nf^3), \\
    \eta_\phi & = 1 + \frac{4(3\pi^2+16)}{3\pi^4 \Nf^2}  + \mathcal O(1/\Nf^3), \\
    \eta_\psi & = \frac{2}{\pi^2 \Nf} + \frac{16}{3\pi^4 \Nf^2}
    + \frac{378 \zeta(3) - 36\pi^2 \ln(2) - 45 \pi^2 - 332}{9\pi^6 \Nf^3} 
    \nonumber \\ & \quad {} 
    + \mathcal O(1/\Nf^4)
\end{align}
where $\zeta(s)$ denotes the Riemann zeta function and $\Nf$ corresponds to the number of four-component Dirac fermions.
On the other hand, for the case of $\Nf = 4$ relevant here, the four-loop exponents in $D=4-\epsilon$ space-time dimensions read in numerical form~\cite{zerf17}
\begin{align}
    1/\nu & = 2-1.4051 \epsilon+0.3018 \epsilon^2-0.3032 \epsilon^3+0.6725 \epsilon^4
    \nonumber \\ & \quad {} + \mathcal O(\epsilon^5) \\
    %
    \eta_\phi & = 0.8889 \epsilon+0.1310 \epsilon^2+0.0136 \epsilon^3+0.0585 \epsilon^4 
    \nonumber \\ & \quad {} + \mathcal O(\epsilon^5) \\
    %
    \eta_\psi & = 0.1667 \epsilon-0.0661 \epsilon^2-0.0697 \epsilon^3+0.0156 \epsilon^4
    \nonumber \\ & \quad {} + \mathcal O(\epsilon^5)
\end{align}
We reemphasize that the number of fermion flavors here is doubled in comparison with the previously much-studied scenario of spin-$1/2$ fermions on the honeycomb monolayer.
Since the series expansions are at best only slowly convergent, we study their Pad\'e approximants, which are defined by
\begin{align}
	[m/n](x) \coloneq \frac{a_0 + a_1 x + \ldots + a_m x^m}{1 + b_1 x + \ldots + b_n x^n},
\end{align}
where $x \in \{\epsilon,1/\Nf\}$ for the $4-\epsilon$ and large-$\Nf$ expansions, respectively, and $m$ and $n$ are nonnegative integers chosen such that $m+n$ agrees with the order to which a particular exponent has been calculated. Here, $n+m = 4$ ($n+m = 2$) for $1/\nu$ and $\eta_\phi$ in the $4-\epsilon$ (large-$\Nf$) expansion, whereas $n+m = 4$ ($n+m = 3$) for $\eta_\psi$.
For a given choice of $m$ and $n$, the coefficients $a_0,\dots,a_m$ and $b_1,\dots,b_n$ are then determined by imposing the boundary condition that the approximant $[m/n]$ produces the correct asymptotics for $x \ll 1$ in agreement with the series expansion.
Evaluating $[m/n]$ at finite values of $x$ yields resummed values of the corresponding observable.

\begin{table}[tbp]
	\renewcommand{\arraystretch}{1.25}
	\caption{Critical exponents of the Gross-Neveu-Heisenberg universality class for $\Nf = 4$ four-component fermion flavors in $D = 2+1$ space-time dimensions, as relevant for the nematic-to-coexistence transition on the honeycomb bilayer. We use results from four-loop $4-\epsilon$ expansion~\cite{zerf17}, second-order $1/\Nf$ expansion (third-order for $\eta_\psi$)~\cite{gracey18}, and functional RG in the improved local potential approximation~\cite{janssen14}.
	Besides the na\"ive extrapolations, we show different Pad\'e approximants $[m/n]$ of the series expansions.
	Those for which it is mathematically impossible to match the original series expansion to all available orders are marked ``n.e.''
	A dash (---) in the entry for an approximant signifies either that sufficient terms are not available in the literature to compute it or that it does not exhaust all the terms available in the literature. 
	$[m/n]_2$ denote two-sided Pad\'e approximants, which take superuniversality relations~\cite{gehring15} into account. Here, we have refrained from showing approximants that exhibit a singularity in $D \in (2,4)$, marked with ``sing.''
	For the functional RG results, we have used two different cutoff schemes, marked as ``linear'' and ``sharp'' in the table.
	}
	\label{tab:exponents-Pade1}
	\begin{tabular*}{\linewidth}{@{\extracolsep{\fill} } l l c c c}
	\hline\hline
	& & $1/\nu$ & $\eta_\phi$ & $\eta_\psi$\\
	\hline
	$1/\Nf$ expansion \cite{gracey18}
	& na\"ive  & $0.96232$ & $1.03902$ & --- \\
	& $[1/1]$ & $0.88829$ & n.e.      & --- \\
	& $[0/2]$ & $0.92700$ & $1.04060$ & n.e. \\
    & na\"ive & --- & --- & $0.05306$ \\
	& $[1/2]$ & --- & --- & $0.05292$ \\
	& $[2/1]$ & --- & --- & $0.05329$ \\
	& $[0/3]$ & --- & --- & n.e.\\
	$4-\epsilon$ expansion \cite{zerf17}
	& na\"ive  & $1.26604$ & $1.09193$ & $0.04654$ \\
	& $[3/1]$ & $0.80250$ & $1.01575$ & $0.04368$ \\
	& $[2/2]$ & $0.79277$ & $1.04180$ & $0.06413$ \\
	& $[1/3]$ & $0.88152$ & $1.11866$ & $0.07337$ \\
	& $[0/4]$ & $0.88841$ & n.e.     & n.e.     \\
	Two-sided Pad\'e
	& $[5/0]_2$ & --- & $1.04988$ & --- \\
	& $[4/1]_2$ & --- & sing. & --- \\
	& $[3/2]_2$ & --- & sing. & --- \\
	& $[2/3]_2$ & --- & sing. & --- \\
	& $[1/4]_2$ & --- & $1.06238$ & --- \\
	& $[0/5]_2$ & --- & n.e. & --- \\
	& $[6/0]_2$ & $0.89489$ & --- & $0.05906$ \\
	& $[5/1]_2$ & $0.83956$ & --- & sing. \\
	& $[4/2]_2$ & sing. & --- & $0.05949$ \\
	& $[3/3]_2$ & sing. & --- & $0.06418$ \\
	& $[2/4]_2$ & $0.84007$ & --- & n.e. \\
	& $[1/5]_2$ & $0.86441$ & --- & n.e. \\
	& $[0/6]_2$ & n.e. & --- & n.e. \\
	Functional RG~\cite{janssen14}
	& linear & $0.87834$ & $1.00929$ & $0.03824$ \\
	& sharp & $0.87187$ & $1.01089$ & $0.03567$\\
	\hline\hline
    \end{tabular*}
\end{table}

The extrapolated values for the present case of $\Nf = 4$ flavors of four-component Dirac fermions in $D=2+1$ space-time dimensions are displayed in Table~\ref{tab:exponents-Pade1}. The spread of all admissible Pad\'e approximants yields a measure of how close to convergence the given series happens to be.
Some Pad\'e approximants cannot mathematically fulfill all the boundary conditions imposed by the asymptotic expansions at the origin. On one hand, this concerns $[0/n]$-type approximants, which cannot describe exponents for which the zeroth-order terms vanish in the series expansion. This applies to $\eta_\psi$ in both $4-\epsilon$ and large-$\Nf$ expansions and $\eta_\phi$ in $4-\epsilon$ expansion, as in all other Gross-Neveu-type universality classes~\cite{janssen18}. On the other hand, the $[1/1]$ approximant cannot satisfy all the boundary conditions for $\eta_\phi$ in large-$\Nf$ expansion, because its $\mathcal{O}(1/\Nf)$ correction happens to vanish, which is a peculiarity of the Gross-Neveu-Heisenberg universality class. All such nonexistent approximants are marked ``n.e.'' in Table~\ref{tab:exponents-Pade1}.

For the $4-\epsilon$ expansion, we can refine the extrapolation by exploiting superuniversality relations near the lower critical space-time dimension of two \cite{gehring15}.
For Gross-Neveu-type universality classes in $2<D<4$ space-time dimensions, we have
\begin{align}
	1/\nu & = (D-2) + \mathcal{O}((D-2)^2), \\
	\eta_\phi & = 2 + \mathcal{O}(D-2), \\
	\eta_\psi & = \mathcal{O}((D-2)^2),
\end{align}
independent of the particular member of the Gross-Neveu family and the number of fermion flavors $\Nf$.
These relations can be used to impose additional boundary conditions at $\epsilon = 2$ on Pad\'e approximants to the $4-\epsilon$ expansion~\cite{janssen14, janssen17b, ihrig18}.
We note that for $\eta_\phi$, only the zeroth-order coefficient in $D-2$ is superuniversal, in contrast to $1/\nu$ and $\eta_\psi$.
The resulting Pad\'e approximants are also shown in Table~\ref{tab:exponents-Pade1}. 
Here, some Pad\'e approximants develop singularities as a function of the expansion parameter, and are hence unreliable as extrapolators; these are marked as ``sing.'' in lieu of any actual numerical value. For $\eta_\phi$, various two-sided Pad\'e approximants turn out to be singular, which may be due to the fact that only the zeroth-order term in $D-2$ is included here.
The refinement using two-sided Pad\'e approximants is especially important for $\eta_\psi$, which is a highly nonmonotonic function of $\epsilon$, vanishing at both $\epsilon = 0$ and $\epsilon = 2$ separately. Such behavior is particularly difficult to capture with a one-sided Pad\'e approximation. We find it satisfying that the estimates from the different two-sided Pad\'e approximations appear overall more stable in comparison with the one-sided approximations.

As a complementary approach to estimate the exponents for $\Nf=4$ and $D=2+1$, we employ the functional RG~\cite{dupuis21}.
To this end, we numerically solve the corresponding fixed-point equations in the improved local potential approximation~\cite{janssen14} for the present case of $\Nf=4$.
We use two different cutoff schemes to assess the stability of our numerical results, namely, a linear cutoff~\cite{litim01}, which satisfies an optimization criterion, as well as a sharp cutoff~\cite{janssen12} for comparison.
The corresponding estimates for $1/\nu$, $\eta_\phi$, and $\eta_\psi$ are displayed in the last two rows of Table~\ref{tab:exponents-Pade1}.
In order to arrive at these estimates, we have employed a simple Taylor expansion of the bosonic effective potential up to 16th order in $\phi$ for the linear cutoff and 20th order for the sharp cutoff.
These orders are chosen such that the numbers displayed in the table are converged within the improved local potential approximation up to the fourth digit after the decimal point.

To arrive at final best-guess estimates combining the results of the three complementary approaches, we employ the two-step averaging procedure outlined in Ref.~\cite{ray21}. The first step is to average over all well-behaved approximations within a given approach. This applies to the existent and nonsingular Pad\'e approximants in the case of the series expansions, and to both employed regulators in the case of the functional RG calculations. 
As the results of the $4-\epsilon$ expansion are included implicitly in the two-sided Pad\'e approximants, we do not incorporate the single-sided Pad\'e approximants in the case of the $4-\epsilon$ expansion.
As for the large-$\Nf$ expansion, we include the na\"ive extrapolation in the average if it is sandwiched by two well-behaved proper approximants $[m/n]$ with $n\geq1$. Note in this context that the two-sided approximants $[5/0]_2$ and $[6/0]_2$ are distinct from the untouched series of hypothetical five-loop and six-loop, respectively, $4 - \epsilon$ expansions, and hence do not count as na\"ive extrapolations in the above sense. 
Having done the ``internal'' average within each method, the second step is to take the mean of the three different averages. We thus arrive at
\begin{align}
	1/\nu & = 0.88(6), &
	\eta_\phi & = 1.035(23), &
	\eta_\psi & = 0.050(12).
    \label{eq:critexp}
\end{align}
In the above, the number in parentheses is the larger of (i)~the spread of the estimates of the three individual approaches and (ii)~the sum of ``internal'' uncertainties within the methods.
The number can hence be understood as a measure of the degree of consistency between the different estimates.
In the case of $1/\nu$ and $\eta_\phi$, we find a particularly good agreement: The uncertainty due to lack of consistency among the three methods is much smaller than the sum of the internal uncertainties. In other words, the three methods ``agree within error bars.'' We note that for the large-$\Nf$ expansion for $\eta_\phi$, the uncertainty in the Pad\'e extrapolation is technically ill-defined, since there exists only one well-defined nontrivial Pad\'e approximant in this case. The internal uncertainty of the large-$\Nf$ estimate for $\eta_\phi$ is hence not included in the final error estimate in Eq.~\eqref{eq:critexp}. However, given that $\Nf$ is quite large and the na\"ively-extrapolated result of the large-$\Nf$ expansion lies quite close to the Pad\'e extrapolated value, the uncertainty due to lack of convergence of the large-$\Nf$ expansion is likely small.
The estimate for $\eta_\psi$ has a larger relative uncertainty, which is likely due to the aforementioned nonmonotonic dependence on the space-time dimension within $D \in (2,4)$, as well as the comparatively small absolute value of the estimate itself.

\section{Conclusions}
\label{sec:conclusions}

In this work, we have studied the competition between nematic and layer-antiferromagnetic orders on the Bernal-stacked honeycomb bilayer.
These two orders appear to be the most promising candidate ground states consistent with experiments in bilayer graphene~\cite{mayorov11, velasco12, freitag12, bao12, veligura12}.
We have demonstrated that these orders generically allow a coexistence phase characterized by both nematicity and antiferromagnetism.
As both signs of nematic~\cite{mayorov11} as well as antiferromagnetic~\cite{velasco12} orders have been reported in low-temperature experiments on different samples, we believe that the actual ground state of bilayer graphene is potentially not too far from the coexistence phase, or may even be within that phase.

We have mapped out the phase diagram of an effective model describing the competition between these two orders and discussed the occurring quantum phase transitions.
The transition between the antiferromagnetic and coexistence orders is weakly first order as a consequence of a cubic term that is symmetry-allowed in the effective potential.
By contrast, the transition between the nematic and coexistence orders turns out to be continuous, and we have identified the corresponding universality class of this quantum critical point.
In particular, we have demonstrated that Lorentz symmetry becomes emergent at this transition in the low-energy limit, despite the fact that the rotational symmetry is spontaneously broken at an intermediate RG stage as a consequence of the background nematic order.
The transition therefore falls into the relativistic Gross-Neveu-Heisenberg quantum universality class, which was previously much studied in the context of the semimetal-to-antiferromagnetic transition on the honeycomb monolayer~\cite{herbut06, herbut09b, janssen14, assaad13, toldin15, otsuka16, buividovich18}.
Consequently, the dynamical critical exponent, describing the relative scaling of time and space in the quantum critical regime, is $z=1$ exactly.
However, for spin-$1/2$ fermions on the honeycomb bilayer, the number of Dirac fermion flavors is doubled in comparison with the spin-$1/2$ realization on the monolayer.
This can be understood as a consequence of the splitting of each of the two inequivalent quadratic band touching points in the noninteracting electron spectrum into two mini-Dirac cones in the nematic state.
The universal exponents characterizing the nematic-to-coexistence quantum critical point on the honeycomb bilayer are therefore generically different from the monolayer situation.
We have exploited previous results that were originally devised in the monolayer context to obtain estimates for the correlation-length exponent $\nu$ and the boson and fermion anomalous dimensions $\eta_\phi$ and $\eta_\psi$ in the present case.
In particular, we have used a four-loop $\epsilon$ expansion around the upper critical dimension~\cite{zerf17}, a second-order large-$\Nf$ expansion (with the fermion anomalous dimension derived at third order)~\cite{gracey18}, and a functional RG approach in the improved local potential approximation~\cite{janssen14}. We have obtained reasonable agreement among the results of these complementary approaches for all exponents calculated.
These predictions may be tested in future numerical simulations of suitable models that realize a nematic-to-coexistence quantum critical point.

In bilayer graphene, the nematic and layer-antiferromagnetic states are very close in energy~\cite{jung11, vafek10a, lemonik10, vafek10b, cvetkovic12, lemonik12}, and the actual low-temperature ground state appears very sensitive to external perturbations.
This suggests the possibility that bilayer graphene could be tuned towards or maybe even through the nematic-to-coexistence quantum phase transition that we have discussed in this work.
The relativistic quantum critical point should then reveal itself in a broad quantum critical regime at finite temperatures, characterized by nontrivial scaling behavior of various observables~\cite{sachdevbook}.
For instance, the real-frequency dynamical spin structure factor should scale in this regime as $\mathcal S(\omega,\vec k) \propto (\omega^2 - c^2 \vec k^2)^{-(2-\eta_\phi)/2}$ with $\eta_\phi \approx 1.0$.
The electronic specific heat should scale as $C_\text{el}(T) \propto T^{d/z}$ with $d=2$ and $z=1$.
Within the coexistence phase, the system develops a full, but anisotropic gap in the electronic spectrum. This should have characteristic consequences for transport experiments: Due to the nematic order in this phase, the electrical conductivity, for instance, should become anisotropic, with a two-fold oscillation as a function of in-plane angle for fixed temperature, but at the same time also exhibit an activated behavior as a function of temperature, arising from the spectral gap.

In this work, we have employed a simple effective model that is expected to capture well the universal aspects of the competition between nematic and antiferromagnetic orders in bilayer graphene. For the future, it would be desirable to identify a realistic microscopic model that allows one to study also nonuniversal aspects of the material. This includes the question of whether signatures of the nematic-to-coexistence quantum critical point should be expected to be readily observable in current experiments. Such an analysis might also reveal possible external parameters that could drive the system towards criticality.

A highly tunable and closely related system that has received significant interest in recent years is twisted bilayer graphene. For certain magic angles between the two honeycomb layers~\cite{bistritzer11}, it shows correlated insulating~\cite{cao18a} or unconventional superconducting~\cite{cao18b} instabilities, depending on the electronic filling. Furthermore, intertwined phases featuring nematic order, potentially also coexisting with superconductivity, have very recently been reported~\cite{cao21}. This suggests that a scenario similar to the one we propose here for Bernal-stacked bilayer graphene may also be relevant for the twisted bilayer configuration. This represents an excellent direction for future investigation.


\begin{acknowledgments}

We are grateful to A.~H.~MacDonald, K.~Novoselov, and A.~Yacoby for very helpful comments on the current status of experiments in bilayer graphene, as well as D.~Stöckinger for some enlightening remarks on power counting in effective field theory and on symmetry-restoring counterterms.
This work has been supported by the Deutsche Forschungsgemeinschaft (DFG) through SFB 1143 (A07, Project No.~247310070), the W\"{u}rzburg-Dresden Cluster of Excellence {\it ct.qmat} (EXC 2147, Project No.~390858490), and the Emmy Noether program (JA2306/4-1, Project No.~411750675).

\end{acknowledgments}

\appendix

\section{Fierz matrix for Gross-Neveu-Heisenberg theory space} \label{app:Fierz}

We record here for completeness the explicit form of the Fierz matrix used to identify a Fierz-complete basis of four-fermion interactions in the Gross-Neveu-Heisenberg theory space [Eq.~\eqref{eq:L-GNH}].
It is given by
\begin{align}
    F = \left(
\begin{array}{@{}*{12}c@{}}
1 & 1 & 1 & 1 & 1 & 1 & 1 & 1 & 1 & 9 & 1 & 1 \\
 -3 & 1 & -1 & -3 & 1 & -1 & 3 & -1 & 3 & 3 & 7 & 3 \\
 -1 & -1 & 1 & -1 & -1 & 1 & 1 & 1 & 1 & 1 & 1 & 9 \\
 3 & -1 & -1 & 3 & -1 & 7 & 3 & -1 & 3 & 3 & -1 & 3 \\
 0 & 0 & -2 & 0 & 8 & -2 & 6 & 2 & -6 & 6 & 2 & -6 \\
 0 & 0 & 2 & 8 & 0 & 2 & 2 & -2 & -2 & 2 & -2 & -2 \\
 -1 & -1 & -1 & 3 & 3 & 3 & 7 & -1 & -1 & 3 & 3 & 3 \\
 3 & -1 & 1 & -9 & 3 & -3 & -3 & 9 & -3 & 9 & -3 & 9 \\
 1 & 1 & -1 & -3 & -3 & 3 & -1 & -1 & 7 & 3 & 3 & 3 \\
 -3 & 1 & 9 & 9 & -3 & -3 & -3 & 1 & -3 & 9 & -3 & 9 \\
 0 & 8 & 2 & 0 & 0 & -6 & -6 & -2 & 6 & 18 & 6 & -18 \\
 8 & 0 & -2 & 0 & 0 & 6 & -2 & 2 & 2 & 6 & -6 & -6 \\
\end{array}
\right),
\end{align}
which acts in the space of four-fermion interactions. More precisely, the Fierz identity reads $F\vec{u} = 0$, where $\vec{u} = \begin{pmatrix}\vec{u}_1 \\ \vec{u}_2\end{pmatrix}$ with
\begin{align}
\vec{u}_1 &= \begin{pmatrix}
[\overline{\psi}_a (\gamma_3 \otimes \sigma_\alpha)\psi_a]^2 + [\overline{\psi}_a (\gamma_5 \otimes \sigma_\alpha)\psi_a]^2 \\
[\overline{\psi}_a (\gamma_\mu \gamma_3 \otimes \sigma_\alpha)\psi_a]^2 + [\overline{\psi}_a (\gamma_\mu \gamma_5 \otimes \sigma_\alpha)\psi_a]^2 \\
[\overline{\psi}_a (\gamma_\mu \gamma_{35} \otimes \sigma_\alpha)\psi_a]^2 \\
[\overline{\psi}_a (\gamma_3 \otimes \mathds{1}_2)\psi_a]^2 + [\overline{\psi}_a (\gamma_5 \otimes \mathds{1}_2)\psi_a]^2 \\
[\overline{\psi}_a (\gamma_\mu \gamma_3 \otimes \mathds{1}_2)\psi_a]^2 + [\overline{\psi}_a (\gamma_\mu \gamma_5 \otimes \mathds{1}_2)\psi_a]^2 \\
[\overline{\psi}_a (\gamma_\mu \gamma_{35} \otimes \mathds{1}_2)\psi_a]^2 \\
\end{pmatrix}, \\
\vec{u}_2 &= \begin{pmatrix}
[\overline{\psi}_a (\mathds{1}_4 \otimes \sigma_\alpha)\psi_a]^2 \\
[\overline{\psi}_a(\gamma_{\mu} \otimes \sigma_\alpha) \psi_a]^2 \\
[\overline{\psi}_a(\gamma_{35} \otimes \sigma_\alpha) \psi_a]^2 \\
(\overline{\psi}_a \psi_a)^2 \\
[\overline{\psi}_a(\gamma_{\mu}\otimes \mathds{1}_2) \psi_a]^2 \\
[\overline{\psi}_a(\gamma_{35} \otimes \mathds{1}_2) \psi_a]^2
\end{pmatrix}.
\end{align}
By Gaussian elimination, we can bring the matrix $F$ into reduced row echelon form. In terms of $\vec{u}_1$ and $\vec{u}_2$, this can be expressed compactly as
\begin{align}
    \vec{u}_1 &= \left(
    \begin{array}{@{}*{6}c@{}}
    1 & 0 & 1 & 2 & 1 & 2 \\
    1 & \frac{1}{2} & -\frac{1}{2} & 2 & \frac{1}{2} & \frac{1}{2} \\
    0 & -\frac{1}{2} & -\frac{1}{2} & -1 & -\frac{1}{2} & -\frac{3}{2} \\
    0 & 1 & 0 & 3 & 0 & 3 \\
    0 & -\frac{1}{2} & \frac{3}{2} & 3 & \frac{3}{2} & -\frac{3}{2} \\
    -1 & \frac{1}{2} & -\frac{1}{2} & 0 & -\frac{3}{2} & -\frac{3}{2} \\
    \end{array}
    \right) \vec{u}_2.
	\label{eq:Fierzred}
\end{align}
Note that in the setup above, interactions present in the four-fermion Lagrangian $\mathcal{L}_\text{GNH}$ [Eq.~\eqref{eq:L-GNH}] are the entries of $\vec{u}_2$ by construction, while those that are not constitute $\vec{u}_1$. Thus, even if a contact interaction not originally included in $\mathcal{L}_\text{GNH}$ is generated during RG flow, it can be rewritten in terms of those that were by using Eq.~\eqref{eq:Fierzred} above.

\begin{widetext}
\section{Evaluation of self-energy diagrams in anisotropic Gross-Neveu-Yukawa-Heisenberg model}
\label{app:selfenerg}

In this appendix, we present details of the derivation of the $\beta$ functions of the Fermi velocities $v_x$ and $v_y$ in the Gross-Neveu-Yukawa-Heisenberg model [Eqs.~\eqref{eq:betavx} and \eqref{eq:betavy}].
As will be manifest shortly, at one loop, the question of whether anisotropy perturbations are relevant or not is independent of the fixed-point values $(h_\star^2,\lambda_\star)$ in Eq.~\eqref{eq:GNY-fixedpoint}.
At this order, it is therefore sufficient to consider the self-energy contributions represented by the diagram in Fig.~\ref{fig:selfenerg}.
The corresponding loop integrals are
\begin{align}
	\text{Fig.~\ref{fig:selfenerg}(a)} &= -h^2 \int_q (\mathds{1}_{2\Nf} \otimes \sigma_\alpha) \langle \psi\overline{\psi}\rangle((1-w)p+q) (\mathds{1}_{2\Nf} \otimes \sigma_\beta) \langle \phi_\alpha \phi_\beta \rangle (wp - q) \equiv \Sigma(p), \\
	\text{Fig.~\ref{fig:selfenerg}(b)} &= h^2 \int_q \operatorname{tr}\left[(\mathds{1}_{2\Nf} \otimes \sigma_\alpha) \langle \psi\overline{\psi}\rangle(q) (\mathds{1}_{2\Nf} \otimes \sigma_\beta) \langle \psi\overline{\psi}\rangle(q + p)\right]\equiv \Pi(p) \delta_{\alpha\beta}, \label{eq:Pi(p)}
\end{align}
where $\Sigma(p)$ and $\Pi(p)$ denote the fermion and boson selfenergies, respectively, with $p =(p_0, p_1, p_2) \equiv (\omega,\vec p)$ as the inflowing 3-momentum in $D=2+1$ space-time dimensions.
We have introduced in the above a momentum-routing parameter $w \in [0,1]$ for the vacuum polarization, because the limit of standard routing ($w \to 0$ or $1$) turns out to be singular in this case. This is another artefact of the regularization scheme that can be resolved by a judicious choice of symmetry-restoring counterterms, as shown below. We note in passing that the definition for the vacuum polarization $\Pi(p)$ in Eq.~\eqref{eq:Pi(p)} is well-defined due to the Pauli matrix relation $\operatorname{tr}(\sigma_\alpha\sigma_\beta) = 2\delta_{\alpha\beta}$.

As mentioned in the main text, to carry out the loop integrals, we first extend the Euclidean time direction to a $(D-2)$-dimensional plane, where $D=4-\epsilon$ is the space-time dimension. The spatial dimension $d=2$ is held fixed, which allows us to deal with the spatial anisotropy in a controlled way.
Before performing the $(D-2)$-dimensional frequency integration, we rescale the momenta as $(v_x q_x, v_y q_y) \mapsto |q_0| \tilde{\vec{q}}$, where $|q_0|$ denotes the radial component of the $(D-2)$-dimensional frequency vector $q_0$.
We thus find for the frequency part of the vacuum polarization
\begin{align}
	\left.\frac{\partial}{\partial p_0^2}\Pi(p)\right|_{p=0}
	&= \frac{4\Nf h^2}{v_x v_y}\int_q \left[\frac{2 \left(2 w^2 - 2w + 1\right)\left(v_x^2 q_x^2 + v_y^2 q_y^2\right)}{\left(q_0^2 + v_x^2 q_x^2 + v_y^2 q_y^2\right)^3}-\frac{3 w^2 - 3 w + 1}{\left(q_0^2 +  v_x^2 q_x^2 + v_y^2 q_y^2\right)^2}\right] \nonumber \\
	&= \frac{\mu^{-\epsilon}}{\epsilon}\frac{4\Nf h^2 S_{2-\epsilon}}{v_x v_y(2\pi)^{2-\epsilon}} \int\!\frac{\rmd^2 \tilde{\vec{q}}}{(2\pi)^2} \left[\frac{\left(2 w^2 - 2w + 1\right)\tilde{\vec{q}}^2}{\left(1 + \tilde{\vec{q}}^2\right)^3}-\frac{3 w^2 - 3 w + 1}{\left(1 + \tilde{\vec{q}}^2\right)^2}\right] 
	\nonumber \\ &
	= \frac{\mu^{-\epsilon}}{\epsilon} \frac{4\Nf h^2 S_{2-\epsilon}}{(2\pi)^{2-\epsilon}} \frac{(1-w)w}{4\pi v_x v_y}.
\end{align}
Analogously, for the momentum part we find
\begin{align}
	\left.\frac{\partial}{\partial p_1^2}\Pi(p)\right|_{p=0} 
	&= \frac{\mu^{-\epsilon}}{\epsilon}\frac{4\Nf h^2 S_{2-\epsilon}}{(2\pi)^{2-\epsilon}} \frac{1}{4\pi}\frac{v_x}{v_y},
	&
	\left.\frac{\partial}{\partial p_2^2}\Pi(p)\right|_{p=0} & 
	= \frac{\mu^{-\epsilon}}{\epsilon}\frac{4\Nf h^2 S_{2-\epsilon}}{(2\pi)^{2-\epsilon}} \frac{1}{4\pi}\frac{v_y}{v_x}.
\end{align}
To evaluate the fermion self-energy, it is useful to introduce the ``master integral''
\begin{align} \label{eq:masterintegral}
	I_{nml}(r,s) & \coloneqq \int_{\mathds{R}^2} \frac{\rmd x \rmd y}{(2\pi)^2} \frac{(x^2)^l}{\left(1 + x^2 + y^2\right)^n \left(1 + r^2 x^2 + s^2 y^2\right)^m}
\end{align}
with $n,m,l \in \mathds{N}$.
In terms of the $I$ functions, we find
\begin{align}
	\left.\frac{1}{4\Nf}\operatorname{tr}\left(\gamma_0\frac{\partial}{\partial \rmi p_0}\Sigma(p)\right)\right|_{p = 0} &= \Nb h^2 \int_q\frac{2 q_0^2}{\left(q_0^2 + q_x^2 + q_y^2\right)^2 \left(q_0^2 + q_x^2 v_x^2+q_y^2 v_y^2\right)} \nonumber\\
	&= \frac{\mu^{-\epsilon}}{\epsilon} \frac{\Nb  h^2 S_{2-\epsilon}}{(2\pi)^{2-\epsilon}} \int \frac{\rmd q_x \rmd q_y}{(2\pi)^2} \frac{2}{\left(1 + q_x^2 + q_y^2\right)^2 \left(1 + q_x^2 v_x^2+q_y^2 v_y^2\right)} \nonumber\\
	&= \frac{\mu^{-\epsilon}}{\epsilon} \frac{2\Nb  h^2 S_{2-\epsilon}}{(2\pi)^{2-\epsilon}} I_{210}(v_x,v_y), \displaybreak[1]\\
	\left.\frac{1}{4\Nf}\operatorname{tr}\left(\gamma_1\frac{\partial}{\partial \rmi p_1}\Sigma(p)\right)\right|_{p = 0} &= \Nb h^2 \int_q \frac{2 v_x q_x^2}{\left(q_0^2 + q_x^2 + q_y^2\right)^2 \left(q_0^2 + v_x^2 q_x^2 + v_y^2 q_y^2\right)} \nonumber\\
	&= \frac{\Nb h^2 S_{2-\epsilon}}{(2\pi)^{2-\epsilon}}\frac{\mu^{-\epsilon}}{\epsilon} (2 v_x)\int \frac{\rmd q_x \rmd q_y}{(2\pi)^2} \frac{q_x^2}{\left(1 + q_x^2 + q_y^2\right)^2 \left(1 + v_x^2 q_x^2 + v_y^2 q_y^2\right)} \nonumber\\
	&= \frac{\mu^{-\epsilon}}{\epsilon}\frac{2\Nb h^2 S_{2-\epsilon}}{(2\pi)^{2-\epsilon}}v_x I_{211}(v_x,v_y), \displaybreak[1]\\
	\left.\frac{1}{4\Nf}\operatorname{tr}\left(\gamma_2\frac{\partial}{\partial \rmi p_2}\Sigma(p)\right)\right|_{p = 0} &= \Nb h^2 \int_q \frac{2 v_y q_y^2}{\left(q_0^2 + q_x^2 + q_y^2\right)^2 \left(q_0^2 + v_x^2 q_x^2 + v_y^2 q_y^2\right)} \nonumber\\
	&= \frac{\mu^{-\epsilon}}{\epsilon}\frac{2\Nb h^2 S_{2-\epsilon}}{(2\pi)^{2-\epsilon}} v_y I_{211}(v_y,v_x),
\end{align}
where we have set $w = 1$ for simplicity, since standard momentum routing is nonsingular in this case.
We have also inserted $\Nb$ as the number of bosonic degrees of freedom, with $\Nb  = 3$ corresponding to the present Heisenberg case. The above equations are valid for general $\Nb$ as long as the generators of the symmetry under which $\phi_\alpha$ transform as a vector commute with the Clifford algebra. Besides the Gross-Neveu-Heisenberg example, this includes the Gross-Neveu-Ising case with $\Nb = 1$, the Gross-Neveu-XY case with $\Nb = 2$, as well as further members of the Gross-Neveu family with $\Nb > 3$~\cite{janssen18}. 

Before extracting renormalization constants from the above results, we need to fix the symmetry-restoring counterterms in Eq.~\eqref{eq:counterterms}. A minimal prescription would be
\begin{align}
	D_\psi &= \lim_{v_x = v_y \to 1} \left[\left.\frac{1}{4\Nf}\operatorname{tr}\left(\gamma_0\frac{\partial}{\partial \rmi p_0}\Sigma(p)\right)\right|_{p = 0} - \left.\frac{1}{4\Nf}\operatorname{tr}\left(\gamma_0\frac{\partial}{\partial \rmi p_0}\Sigma(p)\right)\right|_{p = 0}\right], \\
	D_\phi &= \lim_{v_x = v_y \to 1} \left[\left.\frac{\partial}{\partial p_1^2}\Pi(p)\right|_{p=0} - \left.\frac{\partial}{\partial p_0^2}\Pi(p)\right|_{p=0}\right] \label{eq:symrestphi},
\end{align}
which is precisely what we choose for $D_\psi$. For the bosonic counterterm, we use a slightly modified prescription
\begin{align}
	D_\phi = \frac{D_\phi[\text{Eq.~\eqref{eq:symrestphi}}]}{v_x v_y},
\end{align}
which has the advantage of furthermore canceling all momentum-routing dependence at once (rather than, e.g., at the fixed-point level). We can then read off the remaining renormalization constants in the usual manner. Using $-\mu\rmd(\mu^{-\epsilon}/\epsilon)/\rmd \mu = \mu^{-\epsilon}$, we then arrive at the $\beta$ functions quoted in Eqs.~\eqref{eq:betavx} and \eqref{eq:betavy}.

\section{Master integrals for anisotropic Gross-Neveu-Yukawa-Heisenberg model}
\label{app:mastint}
%
The derivation of the $\beta$ functions of the Fermi velocities $v_x$ and $v_y$ in the anisotropic Gross-Neveu-Yukawa-Heisenberg model involves the master integrals $I_{nml}(v_x,v_y)$ defined in Eq.~\eqref{eq:masterintegral}, more specifically the two functions $I_{210}(v_x,v_y)$ and $I_{211}(v_x,v_y)$. 
These can be evaluated explicitly, and we record the results here for completeness.
For general $v_x, v_y > 0$, they work out to
\begin{align}
    I_{210}(v_x,v_y) &= \frac{\left(v_y^2-1\right) \left(v_x v_y-1\right) \sqrt{\frac{1-v_x^2}{v_y^2-1}}+\left(v_x^2 + v_y^2 - 2 v_x^2 v_y^2\right) \operatorname{arcsin}\left(\sqrt{\frac{v_y^2-1}{v_y^2-v_x^2}}\right) - \left(v_x^2 + v_y^2 - 2 v_x^2 v_y^2\right) \operatorname{arcsin}\left(v_x\sqrt{\frac{v_y^2-1}{v_y^2-v_x^2}}\right)}{4 \pi  \left(1 - v_x^2\right) \left(v_y^2-1\right)^2 \sqrt{\frac{1-v_x^2}{v_y^2-1}}}, \displaybreak[1]\\
	I_{211}(v_x,v_y) &= \frac{1}{8 \pi  \left(v_x^2-1\right) \left(v_y^2-1\right) \left(v_x+v_y\right)
	\sqrt{\frac{1-v_x^2}{v_y^2-1}}} \nonumber \\
	&\phantom{{}={}} \times \left[2 v_x \left(v_y^2-1\right) \sqrt{\tfrac{1-v_x^2}{v_y^2-1}}-3 v_y^2 \left(v_x+v_y\right) \operatorname{arcsin}\left(\sqrt{\tfrac{v_y^2-1}{v_y^2-v_x^2}}\right)+2 \left(v_x+v_y\right) \operatorname{arcsin}\left(v_x\sqrt{\tfrac{v_y^2-1}{v_y^2-v_x^2}}\right) \right. \nonumber\\ 
	&\phantom{{}={}} \qquad\left.+ \left(3 v_y^2-2\right) \left(v_x+v_y\right) \operatorname{arcsin}\left(\sqrt{\tfrac{v_y^2-1}{v_y^2-v_x^2}}\right)\right].
\end{align}
Whenever the argument of a square root obtains a negative value, it is continued analytically as $\sqrt{-a^2} = a\rmi$ for $a \in \mathds{R}_{\geq 0}$. The trigonometric functions are then understood to be replaced by hyperbolic functions in the usual manner. The limits $v_x \to 1$ or $v_y \to 1$ are removable singularities,
\begin{align}
	\lim_{v_x \to 1} I_{210}(v_x,v_y) &= \frac{1 + 2 v_y}{6 \pi (1 + v_y)^2}, &
	\lim_{v_x \to 1} I_{211}(v_x,v_y) &= \frac{1 + 2 v_y}{12 \pi (1 + v_y)^2}, %
	\label{eq:I-vx-1} \\ 
	\lim_{v_y \to 1} I_{210}(v_x,v_y) &= 
	\frac{1 + 2 v_x}{6 \pi (1 + v_x)^2}, &
	\lim_{v_y \to 1} I_{211}(v_x,v_y) &= \frac{1}{4 \pi (1 + v_x)^2}.
    \label{eq:I-vy-1}
\end{align}
In the rotationally-invariant case $v \equiv v_x = v_y$, we obtain the limits
\begin{align}
	I_{210}(v,v) &= \frac{1 - v^2 + 2 v^2 \ln v}{4 \pi  \left(1 - v^2\right)^2}, &
	I_{211}(v,v) &= \frac{v^2 - 2 \ln v - 1}{8 \pi  \left(1 - v^2\right)^2}.
\end{align}
For $v \to 1$, the singularities are again removable, with the pertinent limits given by
\begin{align}
	\lim_{v \to 1} I_{210}(v,v) & = \frac{1}{8\pi}, &
	\lim_{v \to 1} I_{211}(v,v) & = \frac{1}{16\pi},
\end{align}
in agreement with the $v_y \to 1$ and $v_x \to 1$, respectively, limits of Eqs.~\eqref{eq:I-vx-1} and \eqref{eq:I-vy-1}.

\end{widetext}


\bibliographystyle{longapsrev4-2}
\bibliography{bilgra_coexistence}

\end{document}